\documentclass[aip,jcp,reprint,amsmath,amssymb,floatfix,citeautoscript,nofootinbib]{revtex4-2}
\usepackage{times}
\usepackage{mathptmx}
\DeclareMathAlphabet{\mathcal}{OMS}{cmsy}{m}{n} % restore mathcal in mathptmx
\usepackage[T1]{fontenc}
\usepackage{algorithm}
\usepackage{algpseudocode}
\usepackage{bm}
\usepackage{amsmath}
\usepackage{amssymb}
\usepackage{xcolor}
\usepackage{graphicx}
\usepackage{hyperref}
\hypersetup{
    colorlinks=true,
    linkcolor=blue,
    urlcolor=blue,
    citecolor=blue,
}
\usepackage{cleveref}
\usepackage{braket}
\usepackage[version=4]{mhchem}
\usepackage{multirow}
\usepackage{booktabs}

\definecolor{myred}{rgb}{0.8,0,0}
\newcommand{\myred}[1]{{#1}}

\DeclareUnicodeCharacter{2009}{\,}

\begin{document}

    \title{Correlation-consistent Gaussian basis sets for copper solids from material-constrained atomic optimization}

    \author{Jincheng Yu}
    \affiliation{Department of Chemistry and Biochemistry, University of Maryland, College Park, Maryland, 20742, United States}

    \author{Xiaoyu Zhang}
    \affiliation{Department of Chemistry and Biochemistry, University of Maryland, College Park, Maryland, 20742, United States}
    \affiliation{College of Chemistry and Molecular Engineering, Peking University, Beijing, 100871, China}

    \author{Min-Ye Zhang}
    \affiliation{Institute of Physics, Chinese Academy of Sciences, Beijing, 100190, China}
    \affiliation{The NOMAD Laboratory at the Fritz Haber Institute of the Max-Planck-Gesellschaft, Berlin, 14195, Germany}

    \author{Yu Cao}
    \affiliation{School of Physics, Peking University, Beijing, 100871, China}
    \affiliation{Institute of Physics, Chinese Academy of Sciences, Beijing, 100190, China}

    \author{Qiming Sun}
    \affiliation{Quantum Engine LLC, Lacey, Washington 98516, United States}

    \author{Hong-Zhou Ye}
    \email{hzye@umd.edu}
    \affiliation{Department of Chemistry and Biochemistry, University of Maryland, College Park, Maryland, 20742, United States}
    \affiliation{Institute for Physical Science and Technology, University of Maryland, College Park, Maryland, 20742, United States}

    \date{\today}

    \begin{abstract}
        Correlation-consistent Gaussian basis sets enable systematic complete-basis-set (CBS) extrapolation in molecular quantum chemistry, but their diffuse atom-optimized primitives can lead to severe linear dependence in periodic solids, particularly metals and metal-containing compounds.
        Here, we introduce material-constrained atomic optimization (MCAO), which preserves the atomic and correlation-consistent character of Gaussian basis sets while penalizing overlap-matrix ill-conditioning in representative solid-state environments.
        As a proof of concept, we develop Dunning-style MCAO-cc-pV$X$Z ($X = $ D, T, Q) basis sets for Cu with nonrelativistic and scalar-relativistic all-electron potential, effective core potential, and pseudopotential treatments.
        The resulting basis sets exhibit robust numerical stability across bulk Cu, Cu-containing compounds, and Cu surfaces while accurately reproducing plane-wave reference structural properties and electronic band structures.
        Their correlation-consistent construction enables CBS-extrapolated random-phase approximation calculations for bulk Cu and CO adsorption on Cu(111), providing a systematic framework for assessing basis-set, relativistic, and core-treatment errors in correlated electronic-structure calculations of transition-metal solids.
    \end{abstract}

    \maketitle

    \section{Introduction}
    \label{sec:intro}

    Correlated-wavefunction methods are becoming increasingly useful for benchmark calculations of solids and surfaces~\cite{Svelle09JACS,Muller12PCCP,Booth13Nature,Libisch14ACR,McClain17JCTC,Zhang19FM}, where chemically relevant energy differences can depend sensitively on exchange~\cite{Stroppa07PRB,Liu11JCP},
    correlation~\cite{Yang14Science,Masios23PRL,Neufeld23PRL,Ye24JCTC},
    finite-size effects~\cite{Gruber18PRX,Iyer22JPCA,Ye24FD},
    and basis-set convergence~\cite{Olsen13PRB,Morales20JCP,Neufeld22JPCL,Ye22JCTC}.
    These requirements are especially demanding for transition-metal systems, where localized $d$ electrons, metallic screening, and adsorption-induced charge rearrangements must be described on the same footing.~\cite{Hammer95SS}
    The CO adsorption puzzle on transition-metal surfaces~\cite{Feibelman01JPCB,Grinberg02JCP,Patra19PRB} is a representative example: small site-preference energies distinguish the experimentally observed top site from the hollow site favored by semilocal density functional theory (DFT)~\cite{Feibelman01JPCB,Grinberg02JCP,Stroppa08NJP}.
    For such problems, reliable complete-basis-set (CBS) extrapolation is essential for separating electronic-structure errors from basis-set artifacts.

    Gaussian basis sets are attractive for this purpose because molecular quantum chemistry has developed systematic correlation-consistent families that enable reliable CBS extrapolation in both mean-field and correlated-wavefunction calculations.~\cite{Helgaker97JCP,Peterson07ARCC,Hill13IJQC}
    However, standard Gaussian basis sets optimized for isolated atoms or molecules often contain diffuse primitives that become nearly linearly dependent in periodic solids~\cite{Pisani88Book,Peintinger13JCC,Oliveira19JCC,Lee21JCP,Daga20JCTC,Ye22JCTC}.
    Early applications of periodic Gaussian-basis quantum chemistry to solids avoided this issue by using minimal or small split-valence basis sets~\cite{Pisani88Book,Doll00CPL,Neef06SS}, but already noted the potential linear dependence problem of larger basis sets~\cite{Pisani88Book}.
    Over the years, several complementary strategies have been developed and tested, including removal of diffuse exponents~\cite{Geipel97CPL,Paier09PRB,Zhao16JCP,Kadek19PRB,Cao25arXiv}, canonical orthogonalization~\cite{Pisani88Book,Kadek19PRB,Lee21JCP},
    and optimization of exponents and contraction coefficients against representative molecules or solids with constraints on the basis set conditioning~\cite{VandeVondele07JCP,Peintinger13JCC,Oliveira19JCC,Daga20JCTC,Zhou21JCTC,Neufeld22JPCL,Morales20JCP,Li21JPCL}.
    Despite their success in DFT calculations, most existing solid-state basis sets do not have the desired correlation-consistent structure for correlated wavefunction calculations.~\cite{Dunning89JCP}
    One exception is the correlation-consistent GTH-cc-pV$X$Z ($X=$ D, T, Q) basis sets~\cite{Ye22JCTC} developed by one of the authors for the Goedecker-Teter-Hutter (GTH) pseudopotentials~\cite{Hartwigsen98PRB}. These sets have so far been limited to main-group elements in the first two rows of the periodic table.

    In this work, we address the need for solid-state correlation-consistent basis sets by introducing material-constrained atomic optimization (MCAO), which augments standard atomic basis-set optimization with a material-dependent stability constraint.
    The key hypothesis is that the overall structure of an atom-optimized basis set can be largely retained while diffuse exponents are regularized as needed to reduce linear dependence in the chosen material.
    As a proof of concept, we apply MCAO to generate Dunning-style correlation-consistent Cu basis sets, denoted MCAO-cc-pV$X$Z ($X=$ D, T, Q), for a range of nuclear-potential and relativistic treatments.
    The resulting basis sets remain numerically stable for Cu-containing solids while reproducing molecular Cu dimer energetics and plane-wave reference properties of bulk Cu.
    CBS-extrapolated random-phase approximation~\cite{Bohm53PR,GellMann57PR,Furche01PRB,Ren12JMS} (RPA) calculations further enable a controlled assessment of pseudopotential, relativistic, and basis-set errors in bulk Cu and CO adsorption on Cu(111), providing scalar-relativistic all-electron Gaussian-basis benchmarks for bulk Cu and the CO adsorption puzzle.

    \section{Theory}
    \label{sec:theory}

    \subsection{Atomic optimization}
    \label{subsec:atomic_opt}

    Many widely used Gaussian basis sets are generated by atomic optimization,
    \begin{equation}    \label{eq:atom_opt}
        \bm{\alpha}
            \gets \min_{\bm{\alpha}} E_{\text{atom}}(\bm{\alpha})
    \end{equation}
    where $\bm{\alpha} = (\bm{\alpha}^{l = 0}, \bm{\alpha}^{l = 1}, \cdots)$ denotes the primitive Gaussian exponents grouped by angular momentum $l$.
    The atomic objective $E_{\text{atom}}$ may be a mean-field or a correlation energy.
    For example, the correlation-consistent cc-pV$X$Z basis sets introduced by Dunning~\cite{Dunning89JCP} combine a valence set optimized at the HF level with a polarization set optimized at a correlated level, both following \cref{eq:atom_opt} (except for hydrogen and alkali metals, whose polarization functions are optimized using homonuclear diatomic molecules).
    Other prominent examples include the def2 series of Weigend and Ahlrichs~\cite{Weigend03JCP,Weigend05PCCP} and extensions or augmentations designed for spatially extensive electronic states~\cite{Kendall92JCP}, core and semicore correlation~\cite{Woon95JCP}, and relativistic corrections~\cite{Jong01JCP,Pollak17JCTC}.
    In some basis families, such as Jensen's polarization-consistent (pc) sets~\cite{Jensen01JCP,Jensen02JCP}, the polarization functions are optimized at the mean-field level using small elemental molecules.
    Together, these developments provide a rich library of Gaussian basis sets for highly accurate molecular quantum chemistry.~\cite{Hill13IJQC}

    \subsection{The linear dependence issue and prior state of the art}
    \label{subsec:lindep}

\begin{figure}[!t]
        \centering
        \includegraphics[width=3in]{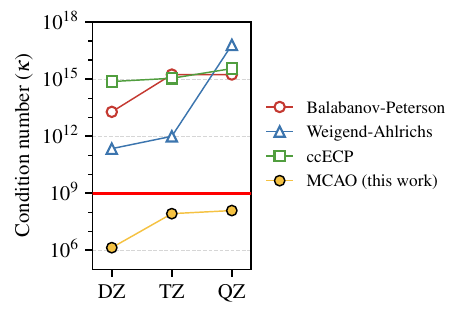}
        \caption{Basis-set condition number $\kappa$ for \textit{fcc} Cu at the equilibrium lattice constant ($3.615~\text{\AA}$) using different Gaussian basis sets: the cc-pVDZ/TZ/QZ sets of Balabanov and Peterson~\cite{Balabanov05JCP}, the def2-SVP/TZVP/QZVP sets of Weigend and Ahlrichs~\cite{Weigend05PCCP}, the ccECP-cc-pVDZ/TZ/QZ sets developed for \myred{the hard Cu ccECP}~\cite{Annaberdiyev18JCP}, and the MCAO-cc-pVDZ/TZ/QZ sets developed in this work.
        The condition number is evaluated from the union of $k$-points in $n \times n \times n$ $\Gamma$-centered meshes with $n = 1, 2, \cdots, 5$.
        }
        \label{fig:cu_cond}
    \end{figure}

    An unconstrained atomic or molecular optimization, however, can generate diffuse basis functions that lead to severe linear dependence in bulk solids and surfaces.~\cite{Pisani88Book,Peintinger13JCC,Ye22JCTC}
    This problem was recognized as early as the 1980s, during the development of periodic Gaussian-basis Hartree-Fock codes.~\cite{Pisani88Book}
    The problem is known to be especially acute for metals and metal-containing solids~\cite{Peintinger13JCC,Ye22JCTC}, where the lower effective nuclear charge of metal atoms often produces very diffuse primitives, for example with $\alpha \ll 0.1~\text{Bohr}^{-2}$.
    \Cref{fig:cu_cond} illustrates this issue for several atom-optimized basis sets of Cu, including the cc-pV$X$Z series~\cite{Balabanov05JCP}, the def2-$X$ZVP series~\cite{Weigend05PCCP}, and the ccECP-cc-pV$X$Z series~\cite{Annaberdiyev18JCP} developed for the correlation-consistent effective core potential~\cite{Bennett17JCP,Annaberdiyev18JCP} (ccECP).
    At DZ and SVP levels, the lowest Gaussian exponents in all three basis-set families are already below $0.04~\text{Bohr}^{-2}$, and basis sets of higher-zeta levels contain even more diffuse basis functions.
    \Cref{fig:cu_cond} characterizes the linear dependence of each basis set by the condition number $\kappa$ of the overlap matrix evaluated for face-centered cubic (\textit{fcc}) Cu.
    Throughout this work, $\kappa$ is defined as the ratio of the maximum to minimum eigenvalues of the overlap matrix evaluated for a sufficiently large supercell of a given lattice.
    Previous work~\cite{Ye22JCTC} has shown that $\kappa \gtrsim 10^{9}$ often causes numerical instabilities, including difficult self-consistent field (SCF) convergence and nonsmooth potential energy surfaces.
    By this criterion, all atom-optimized basis sets considered in \cref{fig:cu_cond} are likely to be unstable for periodic Cu calculations.

    Consistent with this observation, previous Gaussian-basis calculations of bulk Cu or Cu surfaces explicitly avoided the linear dependence issue in different ways.
    Early DFT studies of Cu surface adsorption by Doll and Harrison~\cite{Doll00CPL} and Neef and Doll~\cite{Neef06SS} adopted a small split-valence basis set without polarization functions.
    The authors optimized the outermost $s$, $p$, and $d$ valence exponents using bulk \textit{fcc} Cu and obtained relatively large exponents ($\geq 0.15~\text{Bohr}^{-2}$) that avoided linear dependence.
    More recently, Cao and co-workers~\cite{Cao25arXiv} used the def2-SVP and def2-TZVP basis sets in their correlated wavefunction calculations of Cu surface adsorption.
    They explicitly removed diffuse exponents below $0.05~\text{Bohr}^{-2}$ to avoid linear dependence.
    We will benchmark the accuracy of the resulting truncated basis sets in \cref{sec:results}.
    A similar exponent removal protocol was adopted by Kadek and co-workers~\cite{Kadek19PRB} in an earlier study of bulk silver halide.

    One way to remove linear dependence in a system-specific manner is canonical orthogonalization~\cite{Szabo96Book}, which is employed by default in most quantum chemistry packages.
    In this approach, one diagonalizes the basis set overlap matrix and constructs a basis transformation matrix from eigenvectors whose eigenvalues exceed a user-selected threshold.
    By definition, the eigenvectors being discarded correspond to redundant basis functions that cause linear dependence under the chosen numerical threshold.
    Canonical orthogonalization has been used successfully to stabilize atom-optimized basis sets in solid-state DFT calculations~\cite{Kadek19PRB,Lee21JCP}.
    However, our benchmarks on bulk Cu at RPA level, summarized in figs.~S7 and S8 of the Supporting Information, suggest that correlated-wavefunction calculations can be more sensitive to the numerical threshold used in canonical orthogonalization, leading to discontinuities in the resulting potential energy surfaces.

    In addition to modifying existing atom-optimized basis sets, efforts have also been made to generate numerically stable basis sets through systematic conditioning-aware optimization.
    These include solid-specific~\cite{Daga20JCTC,Morales20JCP,Zhou21JCTC,Neufeld22JPCL} and molecule-specific~\cite{VandeVondele05CPC,VandeVondele07JCP,Li21JPCL} optimizations, in which basis-set parameters are determined by minimizing the energies of representative solids or molecules.
    The resulting basis sets have demonstrated high accuracy and numerical stability in solid-state DFT calculations, including a recent application to Cu surfaces with optimized pseudopotentials~\cite{Li21JPCL}.
    In most cases, however, the atomic optimization protocol and the associated basis-set structure, such as correlation consistency, are either heavily modified or abandoned.
    As a result, the resulting solid-friendly basis sets can have limited transferability~\cite{Daga20JCTC,Morales20JCP,Zhou21JCTC,Erba23JCTC} or may not be suitable for correlated calculations.~\cite{Ye22JCTC}

    One exception is the pob-TZVP basis sets developed by Bredow and co-workers~\cite{Peintinger13JCC}, which are derived from the def2-TZVP sets through controlled solid-specific optimization and have been benchmarked to be portable and numerically stable in mean-field calculations of periodic solids.
    The original pob-TZVP sets were found to have sizable basis set superposition error (BSSE) in certain solids. This issue motivated the pob-TZVP-rev2 sets~\cite{Oliveira19JCC}, which were shown to attain higher accuracy.
    We compare with the pob basis family in \cref{subsec:validation}.

    Another exception is the recently reported GTH-cc-pV$X$Z basis sets by one of the authors optimized for the Goedecker-Teter-Hutter (GTH) pseudopotentials for elements from H to Ar~\cite{Ye22JCTC}.
    These basis sets are atom-optimized and largely preserve the correlation-consistency structure, while controlling linear dependence by judiciously limiting the valence-set size.
    Extensive benchmarks have validated the accuracy of the GTH-cc-pV$X$Z sets for diverse materials and levels of theory~\cite{Ye23arXiv,Ye24JCTC,Ye24FD,Smyser24JPCA,Lau24JPCC,Banerjee22JCTC,Laughon22JPCL,Li24PRL}, highlighting the value of retaining key features from atomic optimization.
    However, the discrete nature of the valence-set selection protocol raises the question of whether accuracy and numerical stability can always be balanced through this route.
    For example, a DZ-quality valence set may be numerically stable but insufficiently accurate, while a TZ-quality valence set may cause severe linear dependence.
    MCAO addresses this problem by enabling nearly continuous tuning of a basis set's degree of linear dependence.

    \subsection{Material-constrained atomic optimization}
    \label{subsec:mcao_theory}

    Inspired by previous work, we introduce material-constrained atomic optimization (MCAO) to generate basis sets that remain stable in solids while closely following their atom-optimized counterparts.
    The design principle is to preserve the standard atomic optimization (\ref{eq:atom_opt}) as much as possible while adding the minimal regularization needed for numerical stability in solids.
    This leads to the following general MCAO loss function for optimizing Gaussian exponents:
    \begin{equation}    \label{eq:mcao_loss}
        \mathcal{L}(\bm{\alpha}; M, \epsilon_0)
            = E_{\text{atom}}(\bm{\alpha})
            + \epsilon_0 P(\bm{\alpha}; M)
    \end{equation}
    where $P(\bm{\alpha}; M)$ is a dimensionless linear-dependence penalty (LDP) evaluated for a reference solid $M$ and specified below.
    The tunable parameter $\varepsilon_0$, which has units of energy, controls the scale at which the LDP competes with the atomic energy $E_{\text{atom}}$ during optimization.
    The MCAO loss function (\ref{eq:mcao_loss}) is formally similar to the one used by VandeVondele and Hutter for generating the molecularly optimized (MOLOPT) basis sets~\cite{VandeVondele05CPC,VandeVondele07JCP}, which was subsequently adapted by others for producing solid-optimized basis sets~\cite{Daga20JCTC,Morales20JCP,Zhou21JCTC,Neufeld22JPCL}.
    An important distinction is that MCAO minimizes the atomic energy, which is independent of the reference material and helps preserve the main structure of atom-optimized basis sets, as shown in \cref{subsec:mcao_cu}.
    This feature places MCAO in the middle ground between standard atomic optimization and prior system-specific approaches, similar in spirit to the design of the pob basis series.~\cite{Peintinger13JCC,Oliveira19JCC}

    In this work, we define the LDP function to be the condition number evaluated on the reference solid,
    \begin{equation}
        P(\bm{\alpha}; M, \kappa_0)
            = \frac{\kappa(\bm{\alpha}; M)}{\kappa_0}
    \end{equation}
    where we have introduced a normalization parameter $\kappa_0$ such that the LDP is effective only when $\kappa \gtrsim \kappa_0$.
    In this sense, $\kappa_0$ plays the role of an approximate target condition number for guiding the MCAO.
    Strictly speaking, only the ratio $\epsilon_0/\kappa_0$ is required to define the MCAO loss \cref{eq:mcao_loss};
    we intentionally retain both parameters to make their distinct roles explicit.
    In practice, we fix $\epsilon_0$ according to the characteristic energy scale of the chosen $E_{\text{atom}}$ (\textit{vide infra}) and tune the strength of the LDP by varying $\kappa_0$.
    In this way, MCAO enables continuous tuning of the conditioning of a basis set on the reference solid, avoiding the coarser control of linear dependence associated with discrete valence-set selection used in previous work~\cite{Ye22JCTC}.
    Other choices of the LDP function are possible~\cite{VandeVondele05CPC}; we leave their exploration for future work.

    \subsection{The MCAO-cc-pVXZ basis sets for Cu}
    \label{subsec:mcao_cu}

    As a proof of concept, we use MCAO to generate correlation-consistent Gaussian basis sets for Cu up to quadruple-zeta (QZ) quality, hereafter denoted MCAO-cc-pV$X$Z with $X$ = D, T, and Q.
    Following established recipes for molecular Cu basis sets~\cite{Weigend05PCCP}, the atomic optimization part of MCAO minimizes the HF energy for the valence set and the MP2 correlation energy for the polarization set;
    we verified that replacing MP2 with coupled-cluster with singles and doubles~\cite{Bartlett07RMP} (CCSD) in the atomic optimization does not significantly alter the optimized polarization exponents.
    Both atomic energies are evaluated for the $[\text{Ar}]3d^{10}4s^{1}$ configuration of Cu, and the MP2 calculation correlates the $11$ valence electrons in the $3d$ and $4s$ shells.
    Because the valence $4p$ shell is unoccupied at the HF level, we follow ref~\onlinecite{Ye22JCTC} and optimize its exponents in the same manner as the polarization functions.

    The MCAO LDP function is evaluated for \textit{fcc} Cu at the experimental lattice constant ($3.615~\text{\AA}$).
    To account for the different energy scales of HF and MP2 correlation energies, we use $\epsilon_0 = 1$, $0.1$, and $0.01$~milli-Hartree for the valence set, the $4p$ shell, and the polarization set, respectively.
    The smaller $\epsilon_0$ for the polarization set helps preserve the correlation-consistency structure, which is essential for the correlated calculations discussed below.
    For each chosen $\epsilon_0$, we initialize MCAO with $\kappa_0$ slightly above the condition number of the input basis and then progressively reduce $\kappa_0$ to a target value of $10^{9}$.
    In standard atomic optimization, the valence and polarization sets are optimized independently and each optimization needs to be performed only once.~\cite{Dunning89JCP,Weigend05PCCP}
    In MCAO, the LDP function couples all exponent subsets, and the subset optimizations are performed iteratively until convergence.
    In practice, three macrocycles are sufficient to converge both the loss function and the basis exponents to reasonable precision.

    After exponent optimization, the valence primitives are contracted using spherically averaged HF atomic natural orbitals (ANOs)~\cite{Widmark90TCA}.
    We then partially decontract the lowest one, two, and three exponents in each angular momentum channel to form the DZ, TZ, and QZ valence sets, respectively.
    We note that HF ANOs yield fewer contracted valence orbitals than the atomic cc-pV$X$Z sets at the same zeta level because the latter use ANOs derived from configuration interaction~\cite{Balabanov05JCP}.
    For Cu, the use of HF ANOs is justified by the closed $3d$ shell in the atomic ground state, which has minimal multireference character.~\cite{Weigend05PCCP}
    For polarization functions, we use a hierarchical $1f$, $2f1g$, and $3f2g1h$ structure for DZ, TZ, and QZ, respectively.
    % The TZ and QZ polarization spaces match those of the atomic cc-pVTZ and cc-pVQZ sets~\cite{Balabanov05JCP}, whereas the atomic cc-pVDZ set uses a $2f \to 1f$ contraction in which the same two $f$ primitives from the TZ set are contracted into one $f$ function.
    % Benchmarks on molecules and solids show no noticeable difference between these two DZ polarization choices.
    % We therefore adopt the simpler $1f$ structure, which also matches the widely used molecular def2-SVP set~\cite{Weigend05PCCP}.
    Combining the valence and polarization sets at each $X$Z level gives the final MCAO-cc-pV$X$Z basis sets.

    We apply this protocol to generate MCAO-cc-pV$X$Z sets for several nuclear-potential and relativistic treatments: the all-electron (AE) potential, hard and soft ccECPs~\cite{Annaberdiyev18JCP,Kincaid22JCP}, and small-core (sc) and large-core (lc) GTH-PBE pseudopotentials~\cite{Hartwigsen98PRB}.
    For the AE potential, we generate basis sets for both nonrelativistic (NR) calculations and calculations with the one-electron spin-free exact-two-component~\cite{Kutzelnigg05JCP,Liu09JCP,Cheng11JCP} (SFX2C-1e) scalar relativistic correction.
    In MCAO, changing the nuclear potential or relativistic treatment primarily affects the atomic energy in \cref{eq:mcao_loss} and the HF ANOs used for contraction, while the solid-state LDP term remains unchanged.
    To account for differences in core size and potential hardness, we adjust the size of the valence primitive set while matching the two or three most diffuse exponents in each angular-momentum channel across nuclear-potential and relativistic treatments.
    A similar exponent matching is achieved automatically for the polarization set by using the same hierarchical structure described above.
    These choices maintain a consistent valence-electron basis-set quality across the different nuclear-potential and relativistic treatments.

    \begin{figure*}[!t]
        \centering
        \includegraphics[width=5.5in]{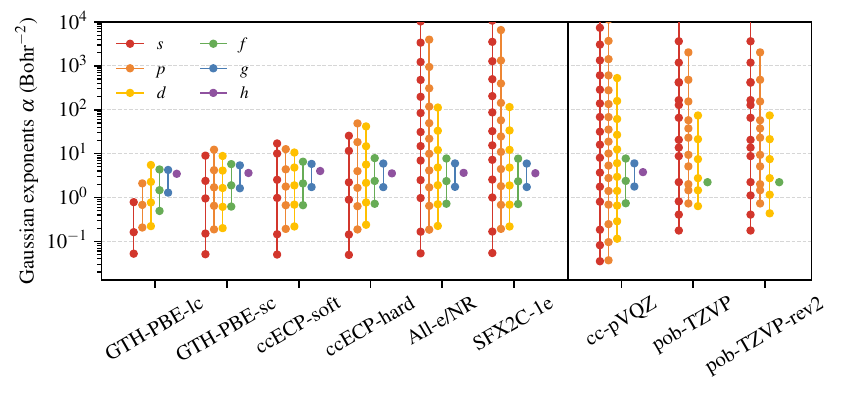}
        \caption{Primitive Gaussian exponents of the Cu MCAO-cc-pVQZ basis sets optimized for different nuclear-potential and relativistic treatments, compared with the standard atomic cc-pVQZ basis set~\cite{Balabanov05JCP}, the pob-TZVP basis set~\cite{Peintinger13JCC}, and the pob-TZVP-rev2 basis set.~\cite{Oliveira19JCC}
        Colors denote angular-momentum channels.
        The valence space contains the $s$, $p$, and $d$ channels, whereas the polarization space contains the $f$, $g$, and $h$ channels.
        The MCAO procedure mainly compacts the most diffuse valence exponents while keeping the polarization exponents close to the atomic correlation-consistent reference.}
        \label{fig:basis_exp_qz}
    \end{figure*}

    The optimized exponents of the MCAO-cc-pVQZ sets for all six nuclear-potential and relativistic treatments are shown in \cref{fig:basis_exp_qz}, illustrating the exponent matching in both the valence and polarization spaces.
    Compared with the standard atomic cc-pVQZ set, the MCAO-cc-pVQZ sets have a more compact valence space, leading to substantially improved numerical stability in solids.
    This improvement is reflected in the reduced $\kappa$ values for the AE/NR MCAO sets in \cref{fig:cu_cond} and for the other nuclear-potential and relativistic treatments in Table~S2.
    Importantly, the polarization exponents of the MCAO sets remain close to those of the atomic cc-pVQZ set, indicating that the correlation-consistency structure is largely preserved.
    \Cref{fig:basis_exp_qz} also shows the exponents of the pob-TZVP and pob-TZVP-rev2 sets for comparison.
    Compared with the MCAO sets, both pob sets have a more compact valence space: their most diffuse valence primitives approximately match the second most diffuse primitives in the MCAO sets.
    The polarization set of both pob sets consists of a single $f$ function, which matches the structure of our MCAO-cc-pVDZ set.
    We discuss the effects of these structural differences on molecular and solid-state calculations in \cref{subsec:validation}.

    \section{Results and discussion}
    \label{sec:results}

    \subsection{Validation of the MCAO-cc-pVXZ sets}
    \label{subsec:validation}

    \subsubsection{Cu dimer}
    \label{subsubsec:cu_dimer}

    \begin{table}[!h]
        \centering
        \caption{All-electron nonrelativistic RPA@PBE atomization energy (eV/atom) of a Cu dimer at a separation of $2.15~\text{\AA}$ calculated with different Gaussian basis sets.
        The def2-TZVP($>0.05$) basis set is derived from def2-TZVP by removing primitives with exponents below $0.05~\text{Bohr}^{-2}$, following ref~\onlinecite{Cao25arXiv}.
        The pob-TZVP+MCAO($s,p,d$) basis set is obtained by augmenting pob-TZVP with the most diffuse $s$, $p$, and $d$ primitives from the AE/NR MCAO-cc-pVTZ basis set.
        }
        \label{tab:cu_dimer}
        \begin{tabular}{lr}
            \toprule
            Basis set & Atomization \\
            \midrule
            cc-pVDZ & $0.70$  \\
            cc-pVTZ & $0.72$  \\
            cc-pVQZ & $0.75$  \\
            CBS(T,Q) & $0.75$  \\
             &   \\
            MCAO-cc-pVDZ & $0.69$  \\
            MCAO-cc-pVTZ & $0.71$  \\
            MCAO-cc-pVQZ & $0.73$  \\
            CBS(T,Q) & $0.74$  \\
             &   \\
            def2-TZVP & $0.69$  \\
            def2-TZVP($>0.05$) & $0.86$  \\
            pob-TZVP & $1.50$  \\
            pob-TZVP-rev2 & $0.83$  \\
            pob-TZVP+MCAO($s,p,d$) & $0.82$  \\
            \bottomrule
        \end{tabular}
    \end{table}

    To examine whether the improved numerical stability of the MCAO basis sets compromises their accuracy, we benchmark molecular and solid-state results against available references.
    For a molecular test, \cref{tab:cu_dimer} reports AE/NR RPA atomization energies of the Cu dimer.
    Throughout this work, all RPA calculations employ Kohn--Sham orbitals obtained with the Perdew--Burke--Ernzerhof (PBE) functional~\cite{Perdew96PRL}; we denote this approach RPA@PBE.
    The MCAO-cc-pV$X$Z results agree with the standard cc-pV$X$Z values within $0.02$~eV/atom (\textit{ca}.~$0.5$~kcal/mol/atom) at each zeta level and in the extrapolated complete-basis-set (CBS) limit.
    This agreement is notable because the MCAO bases are simultaneously constrained to remain numerically stable for Cu solids.

    By contrast, truncating the diffuse exponents in def2-TZVP, as done in ref~\onlinecite{Cao25arXiv} to make the basis set stable for solids, overestimates the atomization energy by $0.17$~eV/atom relative to standard def2-TZVP.
    The latter has a $1f$ polarization set and is therefore expected to have similar quality to cc-pVDZ in correlated calculations, consistent with \cref{tab:cu_dimer}.
    The solid-optimized pob-TZVP set~\cite{Peintinger13JCC} yields an even larger overestimate: its atomization energy is more than twice that obtained with its parent def2-TZVP set.
    This error largely reflects the absence of suitable diffuse valence primitives in pob-TZVP. Augmenting it with the most diffuse $s$, $p$, and $d$ primitives from AE/NR MCAO-cc-pVTZ---denoted pob-TZVP+MCAO($s,p,d$) in \cref{tab:cu_dimer}---reduces the error to $0.10$~eV/atom, comparable to the result from the more recent pob-TZVP-rev2 basis set~\cite{Oliveira19JCC}.
    We emphasize that this simple molecular benchmark should not be interpreted to undermine the value of the pob-TZVP basis sets, which are designed primarily for solid-state calculations.

    \subsubsection{Bulk Cu}
    \label{subsubsec:cu_bulk}

    \begin{figure}[!t]
        \centering
        \includegraphics[width=2.8in]{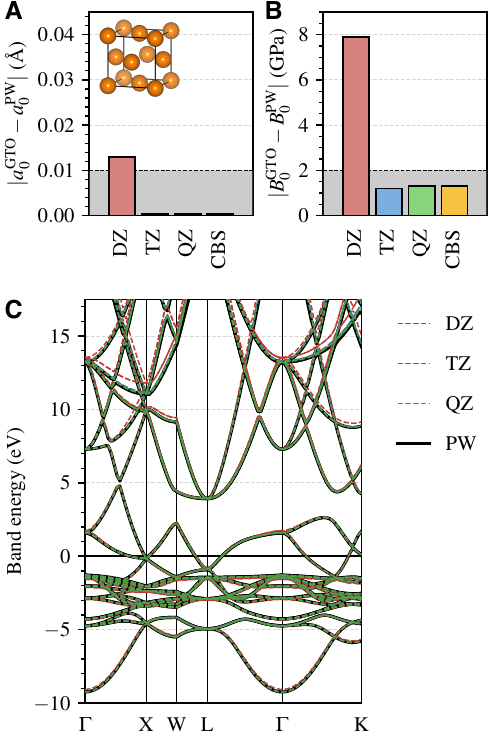}
        \caption{PBE lattice constant, bulk modulus, and band structure of \textit{fcc} Cu calculated with the GTH-PBE-sc pseudopotential and the associated MCAO-cc-pV$X$Z basis sets.
        Panels A and B show the absolute errors of the lattice constant and bulk modulus relative to plane-wave (PW) references obtained with the same pseudopotential.
        The CBS limit is obtained from a (TZ,QZ)-extrapolation.
        The gray shaded area indicates an error within $0.01~\text{\AA}$ and $2$~GPa.
        Panel C compares the MCAO-cc-pV$X$Z and PW PBE band structures, with band energies shifted by the Fermi energy determined in the corresponding basis.
        }
        \label{fig:bulk_basis_conv_pbe}
    \end{figure}

    We next benchmark the MCAO basis sets on periodic solids, beginning with \textit{fcc} Cu, the reference solid used to evaluate the LDP.
    \Cref{fig:bulk_basis_conv_pbe} compares PBE lattice constants ($a_0$), bulk moduli ($B_0$), and band structures of \textit{fcc} Cu calculated with the GTH-PBE-sc pseudopotential.
    With a pseudopotential, a sufficiently converged plane-wave (PW) calculation provides a clear CBS reference.
    The MCAO-cc-pV$X$Z $a_0$ and $B_0$ converge rapidly with zeta quality, reaching the PW reference within $0.01~\text{\AA}$ and $1$~GPa, respectively.
    The high accuracy in ground-state PBE calculations is consistent with an accurate description of the valence bands by the MCAO-cc-pV$X$Z sets.
    In addition, the low-energy conduction bands, which are central to response and correlated-wavefunction calculations, are also reproduced well.
    These results suggest that accurate structural properties and energy spectra can emerge naturally from atom-optimized Gaussian basis sets---once their linear dependence is controlled.
    Explicit solid-specific energy optimization or band-structure fitting, as explored in previous work~\cite{Morales20JCP,Daga20JCTC,Zhou21JCTC,Neufeld22JPCL}, may therefore not be necessary for designing accurate Gaussian basis sets for solids.

    \begin{figure}[!t]
        \centering
        \includegraphics[width=3.4in]{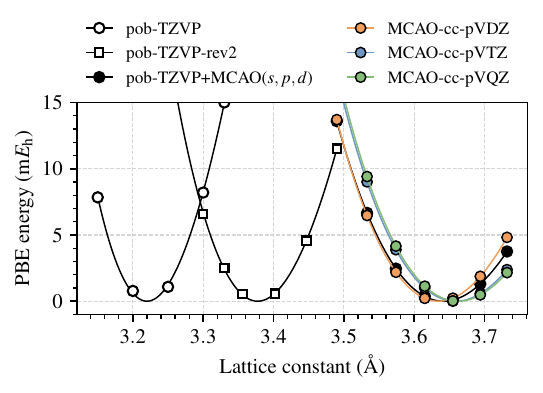}
        \caption{AE/NR PBE energy--lattice constant curves obtained using various basis sets.}
        \label{fig:cmp_pob}
    \end{figure}

    \Cref{fig:cmp_pob} further illustrates the importance of the diffuse valence primitives identified by MCAO by comparing AE/NR PBE energy--lattice constant curves for \textit{fcc} Cu obtained with MCAO-cc-pV$X$Z, pob-TZVP, and pob-TZVP-rev2.
    The MCAO-cc-pV$X$Z curves are qualitatively converged already at the DZ level and have a minimum near $3.65~\text{\AA}$, only slightly above the previously reported PBE value of $3.63~\text{\AA}$~\cite{Harl10PRB}.
    This small difference is primarily due to the neglect of scalar relativistic effects in our AE/NR calculations, as discussed in \cref{subsec:bulk_cu}.
    By contrast, the two pob basis sets predict much more compact lattice structures, underestimating $a_0$ by ca.~$0.4~\text{\AA}$ and $0.3~\text{\AA}$, respectively.
    The curvature of the energy curves predicted by both pob sets is also noticeably larger than that of the MCAO curves, indicating significant overbinding of the Cu lattice.
    This trend is consistent with the overestimated atomization energies obtained with the two pob basis sets in \cref{tab:cu_dimer}.
    As in the molecular dimer benchmark, the errors of the pob basis sets arise primarily from the absence of suitable diffuse valence primitives: the augmented pob-TZVP+MCAO($s,p,d$) basis set gives an energy curve close to the MCAO-cc-pVDZ result (the pob-TZVP polarization set is of DZ quality).
    We emphasize that these comparisons do not undermine the value and established portability of the pob-family basis sets.
    Indeed, augmentation of the pob basis sets with suitable diffuse functions was suggested in the original paper~\cite{Peintinger13JCC}.
    The MCAO algorithm introduced in this work therefore provides a practical and systematic approach to determining the exponents of these important diffuse primitives.

    \subsubsection{Solids of Cu compounds}
    \label{subsubsec:cu_compounds}

    \begin{table}[!h]
        \centering
        \caption{PBE lattice constants ($a_0$) and bulk moduli ($B_0$) calculated for three Cu-compound solids using the MCAO-cc-pV$X$Z basis sets with the GTH-PBE-sc pseudopotential in comparison with plane-wave reference results.}
        \label{tab:cu_compounds}
        \begin{tabular}{lcccccc}
            \hline\hline
            \multirow{2}*{Basis}
                & \multicolumn{3}{c}{$a_0$ ($\text{\AA}$)}
                & \multicolumn{3}{c}{$B_0$ (GPa)}  \\
            \cmidrule(lr){2-4} \cmidrule(lr){5-7}
                & \ce{Cu3N} & \ce{Cu2O} & \ce{CuCl}
                & \ce{Cu3N} & \ce{Cu2O} & \ce{CuCl} \\
            \hline
            MCAO-cc-pVDZ & $3.831$ & $4.314$ & $5.202$ & $115$ & $114$ & $53$ \\
            MCAO-cc-pVTZ & $3.825$ & $4.307$ & $5.169$ & $117$ & $111$ & $57$ \\
            MCAO-cc-pVQZ & $3.823$ & $4.306$ & $5.160$ & $118$ & $112$ & $57$ \\
            Plane wave & $3.823$ & $4.306$ & $5.162$ & $118$ & $112$ & $56$ \\
            \hline
        \end{tabular}
    \end{table}

    To further investigate the transferability of the MCAO basis sets beyond the reference solid, we benchmark their accuracy on solids of three Cu(I) compounds: \ce{Cu3N}, \ce{Cu2O}, and \ce{CuCl}.
    The PBE lattice constants and bulk moduli obtained with the MCAO-cc-pV$X$Z and PW basis sets using the GTH-PBE-sc pseudopotential are shown in \cref{tab:cu_compounds}.
    For the non-Cu elements, we use the GTH-cc-pV$X$Z basis sets from ref~\onlinecite{Ye22JCTC}, \myred{which were originally} optimized for the GTH-HF pseudopotential \myred{but are used here with the GTH-PBE pseudopotential} without modification.
    Both properties converge rapidly with zeta level, as observed for \textit{fcc} Cu.
    For $a_0$, the DZ basis set differs from the PW reference by less than $0.01~\text{\AA}$ for \ce{Cu3N} and \ce{Cu2O}; the TZ and QZ sets reduce the differences to within $0.002~\text{\AA}$.
    For \ce{CuCl}, the DZ difference is larger ($0.04~\text{\AA}$), but the TZ and QZ differences are well below $0.01~\text{\AA}$.
    For $B_0$, the DZ results are within $3$~GPa of the PW references for all three solids, and the TZ and QZ results are within $1$~GPa.
    These results show that the MCAO basis sets retain much of the transferability of atom-optimized basis sets.

    \subsection{Structural properties of bulk Cu}
    \label{subsec:bulk_cu}

    \begin{figure*}[!t]
        \centering
        \includegraphics[width=5.5in]{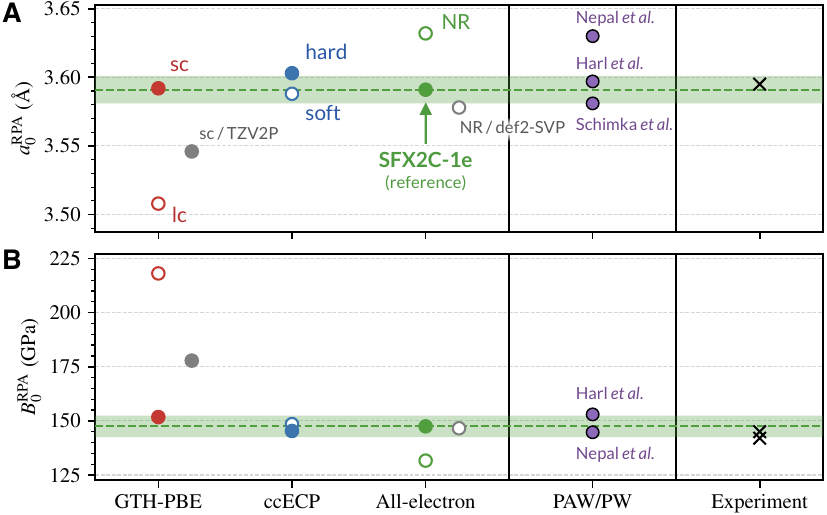}
        \caption{RPA@PBE lattice constant (top) and bulk modulus (bottom) of \textit{fcc} Cu extrapolated to the CBS limit using MCAO-cc-pV(T,Q)Z basis sets for different nuclear-potential and relativistic treatments.
        The SFX2C-1e all-electron result is used as the theoretical reference; the green shaded region indicates deviations within $\pm 0.01~\text{\AA}$ for $a_0$ and $\pm 5$~GPa for $B_0$.
        For the GTH-PBE-sc pseudopotential, the filled gray circle shows our result obtained with the TZV2P-MOLOPT basis set developed by Li and co-workers~\cite{Li21JPCL}.
        For the all-electron potential, the hollow gray circle shows the nonrelativistic def2-SVP result.
        Literature PW/PAW results from Harl \textit{et al.}~\cite{Harl10PRB}, Schimka \textit{et al.}~\cite{Schimka13PRB}, and Nepal \textit{et al.}~\cite{Nepal19PRB}, together with zero-point-energy-corrected experimental references~\cite{Harl10PRB,Schimka11JCP}, are included for comparison.
        }
        \label{fig:rpa_a0b0}
    \end{figure*}

    To demonstrate the practical advantage of the MCAO-cc-pV$X$Z basis sets, we apply them to RPA@PBE equation-of-state calculations for \textit{fcc} Cu.
    Reliable CBS extrapolation with our correlation-consistent MCAO basis sets makes it possible to isolate errors from other approximations, including pseudopotentials, ECPs, and the neglect of relativistic effects.
    Systematic decompositions of basis-set and pseudopotential errors have been explored in the literature primarily at the DFT level.~\cite{Lejaeghere16Science,Bosoni24NRP,Mirhosseini26JCP}
    Here, we use SFX2C-1e all-electron calculations, enabled by recent extensions of SFX2C-1e to periodic solids~\cite{Yeh22PRB}, as the theoretical reference.
    \Cref{fig:rpa_a0b0} compares CBS-extrapolated RPA@PBE $a_0$ and $B_0$ values for bulk Cu obtained with different nuclear-potential and relativistic treatments.
    The GTH-PBE-sc pseudopotential and both ccECPs are in reasonable agreement with the SFX2C-1e all-electron reference.
    In contrast, the GTH-PBE-lc pseudopotential substantially overestimates the Cu--Cu interaction strength, underestimating $a_0$ by approximately $0.1~\text{\AA}$ and overestimating $B_0$ by approximately $70$~GPa.
    This failure cannot be attributed simply to the use of a large core \myred{for two reasons:
    first}, the same number of electrons are correlated in all calculations;
    \myred{second, the PW results discussed below use the same large-core treatment but are in reasonable agreement with our SFX2C-1e reference.}
    The result instead highlights the challenge of parameterizing pseudopotentials for correlated-wavefunction calculations~\cite{Bennett17JCP}.

    \Cref{fig:rpa_a0b0} also demonstrates why reaching the CBS limit is essential for this comparison.
    For the GTH-PBE-sc pseudopotential, we also report our RPA@PBE result obtained with the TZV2P-MOLOPT basis set developed by Li and co-workers~\cite{Li21JPCL} (optimized for the $\omega$B97X-V functional~\cite{Mardirossian14PCCP}) using the MOLOPT approach~\cite{VandeVondele07JCP}.
    Although this basis has been shown to perform well in DFT calculations of Cu-containing solids~\cite{Li21JPCL}, it shows a sizable basis-set incompleteness error in RPA, likely because it lacks a correlation-consistency structure.
    Importantly, this error should not be attributed to the GTH-PBE-sc pseudopotential itself, which is accurate at the CBS limit as demonstrated by our MCAO basis sets.
    The def2-SVP all-electron result provides a similar cautionary example: it appears reasonably accurate relative to the SFX2C-1e reference, but comparison with our CBS-limit nonrelativistic calculation reveals a fortuitous cancellation between basis-set incompleteness and the neglect of scalar relativistic effects.

    Our converged Gaussian-basis SFX2C-1e results, $a_0 \approx 3.59~\text{\AA}$ and $B_0 \approx 147$~GPa, are consistent with independent theoretical benchmarks and experimental references.
    \Cref{fig:rpa_a0b0} includes several literature results obtained with PW bases and the projector augmented wave (PAW) approximation~\cite{Blochl94PRB,Kresse99PRB}, including $a_0$ and $B_0$ from Harl and co-workers~\cite{Harl10PRB} and Nepal and co-workers~\cite{Nepal19PRB}, as well as $a_0$ from Schimka and co-workers~\cite{Schimka13PRB}.
    The PW/PAW framework is widely used for solid-state RPA calculations and has been adopted in recent large-scale benchmarks of RPA-based double-hybrid functionals~\cite{Yu26arXiv,Shi26arXiv}.
    Our best estimate falls within the range of published PW/PAW values.
    It also agrees with zero-point-energy-corrected experimental references compiled by Harl and co-workers~\cite{Harl10PRB} and Schimka and co-workers~\cite{Schimka11JCP}, supporting the accuracy of RPA@PBE for structural properties of transition-metal solids.

    \subsection{CO adsorption on Cu(111) surface}
    \label{subsec:co_ads}

    \begin{figure*}[!t]
        \centering
        \includegraphics[width=6.5in]{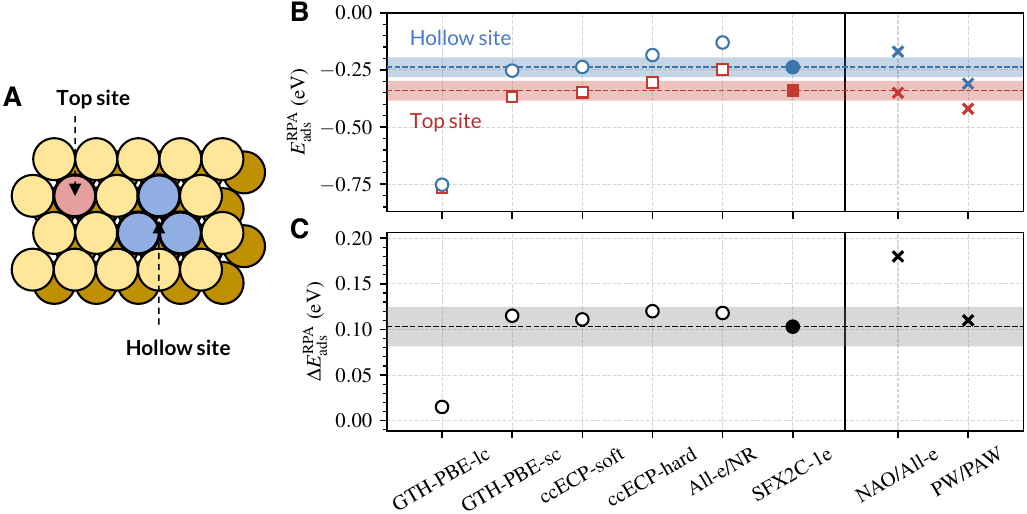}
        \caption{(A) Cu(111) surface model and the top and hollow sites for CO adsorption.
        (B) CBS-extrapolated RPA@PBE adsorption energies of CO on top and hollow sites of Cu(111), obtained using MCAO-cc-pV(T,Q)Z basis sets with counterpoise correction for different nuclear-potential and relativistic treatments.
        The shaded region indicates deviations smaller than $\pm 1$~kcal/mol (\textit{ca}.~$0.04$~eV) from the SFX2C-1e reference.
        Literature results from Ren and co-workers using a numerical atomic orbital (NAO) basis~\cite{Ren09PRB} and from Schimka and co-workers using PW/PAW~\cite{Schimka10NM} are included for comparison.
        (C) RPA@PBE adsorption-energy difference defined in \cref{eq:eads_diff}.
        The shaded region indicates deviations smaller than $\pm 0.5$~kcal/mol (\textit{ca}.~$0.02$~eV) from the SFX2C-1e reference.
        }
        \label{fig:co_ads}
    \end{figure*}

    As a final application, we revisit the CO adsorption puzzle on Cu(111).
    The CO puzzle refers to the failure of semilocal Kohn--Sham DFT to predict the experimentally preferred adsorption site on several transition-metal surfaces~\cite{Feibelman01JPCB,Grinberg02JCP}.
    For Cu(111), shown in \cref{fig:co_ads}A, semilocal DFT favors the threefold \textit{fcc} \textit{hollow} site~\cite{Feibelman01JPCB,Grinberg02JCP,Stroppa08NJP}, largely because of delocalization~\cite{Neef06SS,Stroppa07PRB} or density-driven~\cite{Patra19PRB} errors.
    Low-temperature experiments based on Fourier-transform reflection-absorption infrared spectroscopy~\cite{Raval88SS} (FT-RAIRS) and angle-resolved photoemission extended fine structure~\cite{Moler96PRB} (ARPEFS), however, observe only signatures of adsorption at the onefold \textit{top} site.
    Early studies by Ren and co-workers~\cite{Ren09PRB} and Schimka and co-workers~\cite{Schimka10NM} showed that RPA based on a semilocal DFT reference recovers the correct site preference.
    More recent calculations beyond RPA, including quantum Monte Carlo~\cite{Hsing19JPCC,Iyer22JPCA}, coupled cluster~\cite{Carbone24FD}, and multireference configuration interaction~\cite{Sharifzadeh08JPCC}, have confirmed the RPA site preference and further refined the adsorption energies relative to experiment.

    Most previous periodic calculations used PW bases with pseudopotentials~\cite{Stroppa08NJP,Schimka10NM,Carbone24FD,Hsing19JPCC}, whereas atom-centered local bases have mainly been explored through cluster embedding approximations~\cite{Sharifzadeh08JPCC,Ren09PRB}.
    Cao and co-workers recently reported fully periodic Gaussian-basis RPA@PBE calculations of CO@Cu(111) adsorption energies~\cite{Cao25arXiv}.
    Although those calculations used relatively small modified def2-SVP and def2-TZVP basis sets and neglected relativistic effects, their RPA@PBE adsorption energies were in reasonable agreement with previous PW results.
    Here, we revisit this problem using MCAO-cc-pV$X$Z basis sets to remove basis-set incompleteness in a controlled way.
    This allows us to examine, on a common CBS-limit footing, the effects of pseudopotentials, ECPs, and scalar relativistic corrections.

    \begin{table}[!h]
        \centering
        \caption{\myred{CBS-extrapolated RPA@PBE adsorption energies for CO on the top and hollow sites of Cu(111) and the corresponding site-preference energies, obtained with MCAO-cc-pV$X$Z basis sets for different pseudopotential, ECP, and scalar relativistic treatments.
        Literature results obtained with NAO and PW basis sets are included for comparison.
        All energies are reported in eV.}
        }
        \label{tab:co_ads}
        \begin{tabular}{lccc}
            \toprule
            Method
                & $\Delta E_{\text{ads}}^{\text{top}}$
                & $\Delta E_{\text{ads}}^{\text{hollow}}$
                & $\Delta E_{\text{ads}}$ \\
            \midrule
            GTH-PBE-lc & $-0.77$ & $-0.75$ & $0.01$  \\
            GTH-PBE-sc & $-0.37$ & $-0.25$ & $0.12$  \\
            ccECP-soft & $-0.35$ & $-0.24$ & $0.11$  \\
            ccECP-hard & $-0.30$ & $-0.18$ & $0.12$  \\
            All-e/NR & $-0.25$ & $-0.13$ & $0.12$  \\
            All-e/SFX2C-1e & $-0.34$ & $-0.24$ & $0.10$  \\
            & & &   \\
            All-e/NAO~\cite{Ren09PRB}
                & $-0.35$ & $-0.17$ & $0.18$  \\
            PAW/PW~\cite{Schimka10NM}
                & $-0.42$ & $-0.31$ & $0.11$   \\
            \bottomrule
        \end{tabular}
    \end{table}

    \myred{\Cref{tab:co_ads}} reports CBS-extrapolated RPA@PBE adsorption energies for CO on the top and hollow sites using different nuclear-potential and relativistic treatments.
    \myred{These results are shown in} \cref{fig:co_ads}B.
    All reported adsorption energies include a counterpoise correction for basis-set superposition error~\cite{Boys70MP}. This correction follows established protocols~\cite{Shi23JACS,Ye23arXiv,Ye24FD,Shi25CPC} and is detailed in the Supporting Information.
    \myred{To reduce finite-size errors, we shift all $k$-point meshes by an appropriate mean-value point~\cite{Baldereschi73PRB} and use a hierarchical composite correction to approach the combined basis-set and thermodynamic limits.
    Detailed finite-size convergence studies are presented in the Supporting Information}.

    Relative to the all-electron SFX2C-1e reference, the GTH-PBE-sc pseudopotential and the soft ccECP give adsorption energies within $0.05$~eV for both sites.
    The hard ccECP underestimates both adsorption energies by \textit{ca}.~$0.05$~eV, while nonrelativistic all-electron calculations underestimate them by \textit{ca}.~$0.1$~eV.
    These errors are largely systematic across the two adsorption sites and therefore cancel in the site-preference energy,
    \begin{equation}    \label{eq:eads_diff}
        \Delta E_{\text{ads}}
            = E_{\text{ads}}^{\text{hollow}} - E_{\text{ads}}^{\text{top}}
    \end{equation}
    shown in \cref{fig:co_ads}C.
    Except for GTH-PBE-lc, all pseudopotential, ECP, and nonrelativistic all-electron treatments reproduce the SFX2C-1e $\Delta E_{\text{ads}}$ within $0.02$~eV (\textit{ca}.~$0.5$~kcal/mol).
    The GTH-PBE-lc pseudopotential is the clear outlier: it overbinds CO at both sites by more than $0.4$~eV and predicts an adsorption-energy gap close to zero.
    These adsorption-energy trends mirror the bulk structural trends in \cref{fig:rpa_a0b0}, suggesting that the errors associated with different nuclear potentials and relativistic treatments are systematic and transferable between bulk Cu and Cu surfaces.

    \Cref{fig:co_ads}(B,C) also compares our results with previous RPA@PBE calculations by Ren and co-workers using an all-electron numerical atomic orbital (NAO) basis~\cite{Ren09PRB} and by Schimka and co-workers using PW/PAW~\cite{Schimka10NM}.
    Our SFX2C-1e adsorption energies are in reasonable agreement with the NAO results of Ren and co-workers, while our site-preference energy agrees more closely with the PW/PAW result.
    Because all three studies made substantial efforts to approach the CBS limit, the remaining differences likely arise from other sources, including \myred{residual finite-size errors}, relativistic treatments, slab models, and pseudopotential approximations.
    Resolving these differences will require coordinated comparisons across codes, which we leave for future work.
    The MCAO-cc-pV$X$Z sets developed here provide a controlled Gaussian-basis route for such comparisons because the basis-set error can be systematically reduced while maintaining numerical stability.

    We also comment on the fully periodic Gaussian-basis RPA@PBE calculations of Cao and co-workers, who reported adsorption energies of $-0.31$ and $-0.19$~eV for the top and hollow sites, respectively, corresponding to a site-preference energy of $0.11$~eV~\cite{Cao25arXiv}.
    These values are in reasonable agreement with our SFX2C-1e reference and with the PW/PAW results of Schimka and co-workers.~\cite{Schimka10NM}
    However, we found that this comparison is sensitive to both basis-set and finite-size extrapolation.
    As discussed in the Supporting Information, our calculations with the same modified def2-SVP and def2-TZVP basis sets give substantially less favorable adsorption energies after CBS(SVP,TZVP) extrapolation. An analogous CBS(DZ,TZ) extrapolation with nonrelativistic all-electron MCAO-cc-pV$X$Z basis sets gives similar values.
    This consistency suggests that the remaining discrepancy is primarily associated with the absence of QZ-quality information in a DZ/TZ-only extrapolation, together with differences in the treatment of finite-size effects, rather than with the use of Gaussian basis sets per se.

    \section{Conclusion}
    \label{sec:conclusion}

    To conclude, we have introduced material-constrained atomic optimization as an effective strategy for generating Gaussian basis sets that retain the accuracy of atomic optimization while remaining numerically stable for periodic solids.
    We applied MCAO to construct correlation-consistent Cu basis sets for several pseudopotentials, ECPs, and all-electron potentials, including nonrelativistic and SFX2C-1e scalar relativistic treatments.
    The resulting MCAO-cc-pV$X$Z basis sets reproduce Cu dimer binding energies from standard cc-pV$X$Z sets and accurately recover PW reference structural properties and band structures for bulk Cu.
    The accuracy also extends to Cu-containing solids that were not used as references in the basis-set optimization.
    The reliable CBS extrapolation of the MCAO-cc-pV$X$Z sets enables a controlled assessment of pseudopotential, ECP, and relativistic errors in RPA@PBE calculations of bulk Cu.
    We further applied these basis sets to CO adsorption on Cu(111), obtaining scalar-relativistic all-electron RPA@PBE benchmarks for the top and hollow adsorption sites.

    The present MCAO formulation has several limitations that point to directions for future development.
    First, basis-set linear dependence is often associated with multiple ill-conditioned modes of the overlap matrix, whereas the $\kappa$-based LDP used here depends only on the most extreme eigenvalues.
    A more flexible regularization could penalize all near-singular modes, for example through a smooth threshold applied to the small-eigenvalue spectrum.
    Second, the current MCAO scheme controls linear dependence only through the primitive exponents.
    Future work could extend the optimization to the contraction coefficients, an avenue already exploited in the MOLOPT basis sets to improve numerical stability~\cite{VandeVondele07JCP,Li21JPCL}.
    Third, extending MCAO beyond Cu to other transition-metal and main-group solids may require element-specific refinements, because both the atomic electronic structure and the solid-state bonding environment can change the balance between accuracy and stability.
    Nevertheless, the central idea of combining atomic accuracy with material-dependent stability constraints should provide a useful foundation for future solid-state Gaussian basis-set optimization.

    \section*{Supporting Information}

    The Supporting Information contains (i) geometry and basis-set data files; (ii) computational details; (iii) additional details of MCAO basis-set generation and construction of the corresponding auxiliary basis sets for density fitting; (iv) convergence of the bulk Cu and CO/Cu(111) calculations to the CBS and thermodynamic limits; (v) a comparison with the modified def2 basis sets for CO/Cu(111) adsorption; and (vi) Cu RPA@PBE equation-of-state calculations with canonical orthogonalization.

    \section*{Conflict of interest}
    The authors declare no competing financial interest.

    \section*{Data availability}

    The data that support the findings of this study are available in the GitHub repository: \href{https://github.com/hongzhouye/supporting\_data/tree/main/2026/MCAO\_Cu}{https://github.com/hongzhouye/supporting\_data/tree/main/2026/MCAO\_Cu}.

    \section*{Acknowledgments}

    This work was supported by the National Science Foundation under Grant No.~CHE-2543461.
    We acknowledge computing resources provided by the Division of Information Technology at the University of Maryland, College Park.

    \bibliography{refs}

\end{document}

\title{Supporting Information for Correlation-consistent Gaussian basis sets for copper solids from material-constrained atomic optimization}

    \author{Jincheng Yu}
    \affiliation{Department of Chemistry and Biochemistry, University of Maryland, College Park, Maryland, 20742, United States}

    \author{Xiaoyu Zhang}
    \affiliation{Department of Chemistry and Biochemistry, University of Maryland, College Park, Maryland, 20742, United States}
    \affiliation{College of Chemistry and Molecular Engineering, Peking University, Beijing, 100871, China}

    \author{Min-Ye Zhang}
    \affiliation{Institute of Physics, Chinese Academy of Sciences, Beijing, 100190, China}
    \affiliation{The NOMAD Laboratory at the Fritz Haber Institute of the Max-Planck-Gesellschaft, Berlin, 14195, Germany}

    \author{Yu Cao}
    \affiliation{School of Physics, Peking University, Beijing, 100871, China}
    \affiliation{Institute of Physics, Chinese Academy of Sciences, Beijing, 100190, China}

    \author{Qiming Sun}
    \affiliation{Quantum Engine LLC, Lacey, Washington 98516, United States}

    \author{Hong-Zhou Ye}
    \email{hzye@umd.edu}
    \affiliation{Department of Chemistry and Biochemistry, University of Maryland, College Park, Maryland, 20742, United States}
    \affiliation{Institute for Physical Science and Technology, University of Maryland, College Park, Maryland, 20742, United States}

    \maketitle

    \tableofcontents

    \clearpage

    \section{Geometry and basis-set data}

    Geometry files for the CO/Cu(111) adsorption systems and data files for all Gaussian basis sets used in this work are available in the following GitHub repository:
    \begin{center}
        \href{https://github.com/hongzhouye/supporting\_data/tree/main/2026/MCAO\_Cu}{https://github.com/hongzhouye/supporting\_data/tree/main/2026/MCAO\_Cu}
    \end{center}
    The basis-set data include
    \begin{itemize}
        \item the MCAO-cc-pV$X$Z sets optimized in this work;
        \item the truncated def2-SVP and def2-TZVP sets generated following the protocol of ref~\citenum{Cao25arXiv}; and
        \item the TZV2P-MOLOPT basis sets from ref~\citenum{Li21JPCL}.
    \end{itemize}
    For each orbital basis set, the repository also provides a corresponding auxiliary basis set optimized for density fitting using the protocol described in \cref{sec:aux_basis}.

    \section{Computational details}

    This section summarizes the computational settings used throughout this work.
    Details for basis-set generation, auxiliary basis-set optimization, bulk calculations, and surface calculations are given in \cref{sec:mcao_basis,sec:aux_basis,sec:bulk,sec:surf}.

    \subsection{PySCF}

    All Gaussian-based calculations are performed using the PySCF code~\cite{Sun18WIRCMS,Sun20JCP}.
    The RPA total energy is the sum of three terms, all evaluated using PBE orbitals,
    \begin{equation}
        E^{\text{RPA}}
            = E^{\text{HF}} + \Delta E^{\text{ACFD-X}} + E^{\text{RPA}}_{\text{corr}}
    \end{equation}
    where $E^{\text{HF}}$ is the HF energy, $\Delta E^{\text{ACFD-X}}$ is the difference between the adiabatic-connection fluctuation-dissipation (ACFD) exchange energy and the HF exchange energy discussed in ref~\citenum{Harl10PRB}, and $E^{\text{RPA}}_{\text{corr}}$ is the RPA correlation energy evaluated within the ACFD approach using $40$ quadrature points for the imaginary-frequency integration.
    For the molecular dimer RPA calculations, the PBE reference is calculated at zero temperature without smearing, and the $\Delta E^{\text{ACFD-X}}$ term vanishes.
    For periodic bulk Cu and Cu surface RPA calculations, the PBE reference is calculated with the Fermi-Dirac smearing with a width of $0.01$~eV, so the $\Delta E^{\text{ACFD-X}}$ term is nonzero whenever fractional occupations are present.
    For all nuclear potentials, the correlation energy is calculated with the [Ar] core of Cu and the [He] cores of C and O frozen wherever these electrons are explicitly present; this choice is consistent with most previous RPA calculations on bulk Cu and CO/Cu(111).
    For both $E^{\text{HF}}$ and $\Delta E^{\text{ACFD-X}}$, the integrable divergence in the electron-repulsion integrals is handled using the Ewald probe-charge approach~\cite{Paier05JCP}, which amounts to shifting the total energy by $\frac{n_{\text{e}} v_{\text{M}}}{2}$, where $n_{\text{e}}$ is the total number of electrons per cell and $v_{\text{M}}$ is the Madelung constant~\cite{Broqvist09PRB}.
    Density fitting is used for both molecular and periodic calculations to approximate the electron repulsion integrals.~\cite{Ye21JCPa,Ye21JCPb}
    Auxiliary basis sets are optimized for the MCAO-cc-pV$X$Z orbital basis sets as described in \cref{sec:aux_basis}.

    \subsection{Quantum Espresso}

    Quantum Espresso (QE) is used for the plane-wave (PW) reference calculations on \textit{fcc} Cu with the four pseudopotentials/ECPs summarized in \cref{tab:qe_pp}.
    For the GTH-PBE-lc pseudopotential and the two ccECPs, the Unified Pseudopotential Format (UPF) files are taken directly from QE's pseudopotential library (for GTH-PBE-lc) or from the QMCPACK \texttt{pseudopotentiallibrary} GitHub repository (for the two ccECPs).
    For the GTH-PBE-sc pseudopotential, we generate the UPF file directly from the pseudopotential parameters using the open-source \texttt{gth2upf} library developed by one of the authors~\cite{gth2upf}.
    We verified for GTH-PBE-lc that the standard UPF file and the UPF file generated by \texttt{gth2upf} give identical results.
    For each pseudopotential/ECP, the kinetic-energy cutoff is chosen sufficiently large to converge the lattice constant and bulk modulus to better than $0.001$~\text{\AA} and $0.1$~GPa, respectively, which is adequate for benchmarking the Gaussian-basis results.

    \begin{table}[!h]
        \centering
        \caption{Pseudopotentials and ECPs used for the QE calculations.}
        \label{tab:qe_pp}
        \begin{tabular}{lccc}
            \toprule
            Pseudopotential/ECP & UPF file & Link & \texttt{ecutwfc} \\
            \midrule
            GTH-PBE-lc
                & Cu.pbe-d-hgh.UPF
                & \href{https://pseudopotentials.quantum-espresso.org/upf\_files/Cu.pbe-d-hgh.UPF}{link}
                & $300$~Ry  \\
            GTH-PBE-sc
                & Converted by \texttt{gth2upf}
                & \href{https://github.com/maki49/gth2upf}{link}
                & $600$~Ry  \\
            ccECP-soft
                & Cu.ccECP-soft.upf
                & \href{https://github.com/QMCPACK/pseudopotentiallibrary/blob/main/recipes/Cu/ccECP-soft/Cu.ccECP-soft.upf}{link}
                & $300$~Ry  \\
            ccECP-hard
                & Cu.ccECP.upf\_deprecated
                & \href{https://github.com/QMCPACK/pseudopotentiallibrary/blob/main/recipes/Cu/ccECP/Cu.ccECP.upf\_deprecated}{link}
                & $1000$~Ry \\
            \bottomrule
        \end{tabular}
    \end{table}

    \subsection{VASP}

    VASP PBE calculations are used for bulk Cu and CO/Cu(111) to assess convergence to the thermodynamic limit (TDL).
    The single-point calculations use the \texttt{Cu\_sv\_GW}, \texttt{C\_h}, and \texttt{O\_h} POTCAR files for Cu, C, and O, respectively, with \texttt{ENCUT} set to $1000$~eV.
    For CO/Cu(111), VASP is also used to optimize the geometry at the PBE level.
    These geometry optimizations use the default \texttt{Cu}, \texttt{C}, and \texttt{O} POTCAR files with \texttt{ENCUT} set to $500$~eV.

    \section{MCAO-cc-pV$X$Z basis sets}
    \label{sec:mcao_basis}

    The construction of the MCAO-cc-pV$X$Z basis sets for a given valence-set structure is outlined in \cref{subsec:atom_opt,subsec:mcao}, while the selection of the valence-set structure is discussed in \cref{subsec:val_select}.

    \subsection{Atomic optimization}
    \label{subsec:atom_opt}

    We first perform atomic optimization to generate standard cc-pV$X$Z basis sets, which serve as the initial guess for MCAO.
    For Cu, this optimization proceeds in three steps that minimize either the restricted open-shell HF (ROHF) energy or the unrestricted MP2 correlation energy evaluated from the ROHF reference.
    Both energies are evaluated for the [Ar]$3d^{10}4s^{1}$ ground state of Cu, and the correlation energy is calculated with the [Ar] core kept frozen.
    \begin{enumerate}
        \item The atomic ROHF energy is minimized to determine the exponents of primitives that are occupied in the ROHF ground state.
        \item The MP2 correlation energy is minimized to determine the exponents of primitives for the valence $4p$ shell, which is unoccupied in the ROHF ground state.
        The exponents optimized from step 1 are kept fixed during this step.
        \item The MP2 correlation energy is minimized to determine the exponents of the polarization set, with the valence set exponents from the first two steps kept fixed.
    \end{enumerate}
    After all exponents are optimized, the valence set (i.e.,~the exponents optimized in steps 1 and 2) is contracted using the coefficients of atomic natural orbitals (ANOs) calculated at the ROHF level,
    \begin{equation}
        \phi_{l \mu m}(\bm{r})
            = \sum_{i=1}^{N_{l}} \chi_{l i m}(\bm{r}) C_{l i \mu}
    \end{equation}
    where $\mathbf{C}_l$ is an $N_{l} \times n_{l}$ coefficient matrix that contracts $N_l$ primitive basis functions $\{\chi_{lim}\}$ within the angular-momentum channel $l$ into $n_{l}$ contracted basis functions $\{\phi_{l\mu m}\}$ in the same channel.
    The contraction coefficients are determined by solving the generalized eigenvalue equation
    \begin{equation}    \label{eq:ANO_eqn}
        \mathbf{s}_l \mathbf{d}_l \mathbf{s}_l \mathbf{C}_l
            = \mathbf{s}_l \mathbf{C}_l \bm{\Lambda}_l
    \end{equation}
    and retaining eigenvectors $C_{l i\mu}$ with nonzero eigenvalues $\lambda_{l \mu}$.
    In \cref{eq:ANO_eqn}, $\mathbf{s}_l$ and $\mathbf{d}_l$ are the spherically averaged radial parts of the atomic overlap and density matrices, respectively,
    \begin{equation}
    \begin{split}
        s_{lij}
            &= \frac{1}{2l+1} \sum_{m=-l}^{l} S_{lim,ljm},  \\
        d_{lij}
            &= \frac{1}{2l+1} \sum_{m=-l}^{l} D_{lim,ljm},  \\
    \end{split}
    \end{equation}
    This ANO approach determines the number of contracted GTOs in each shell for the chosen nuclear potential.
    For example, four, two, and one $s$, $p$, and $d$-type orbitals are generated automatically for all-electron basis sets, while the numbers become two, one, and one for small-core pseudopotentials/ECPs with $19$ valence electrons (i.e.,~$3s^2 3p^6 3d^{10} 4s^1$).
    The most diffuse one, two, and three primitives in the $s$, $p$, and $d$ shells are then decontracted from the fully contracted valence sets to make the final DZ, TZ, and QZ valence sets.
    For the $p$ shell, the primitive optimized for the unoccupied $4p$ shell is retained outside the ROHF ANO contraction, compensating for the absence of an occupied $4p$ ANO at the HF level.
    For the polarization set, we use $1f$, $2f1g$, and $3f2g1h$ for DZ, TZ, and QZ, respectively, which is a natural transition-metal analogue of the $1d$, $2d1f$, and $3d2f1g$ hierarchy for main-group elements.
    As discussed in the main text, our TZ and QZ polarization sets have the same structure as the molecular cc-pVTZ and cc-pVQZ basis sets, while our DZ is similar to that of def2-SVP.

    \subsection{MCAO}
    \label{subsec:mcao}

    Starting from an atomically optimized basis set, MCAO follows the same three-step protocol, but each step includes a condition-number penalty evaluated with the full basis set at the current exponents.
    For Cu in particular, we have
    \begin{equation}    \label{eq:mcao_loss_three_term}
    \begin{split}
        \bm{\alpha}_{\text{val}}^{(n+1)}
            &\gets \min_{\bm{\alpha}_{\text{val}}}
            E_{\text{atom}}^{\text{HF}}(\bm{\alpha}_{\text{val}})
            + \frac{\varepsilon_{0,\text{val}}}{\kappa_0} \times
            \max_{M \in \{M\}} \kappa(
                \bm{\alpha}_{\text{val}},
                \bm{\alpha}_{4p}^{(n)},
                \bm{\alpha}_{\text{pol}}^{(n)}; M
            )   \\
        \bm{\alpha}_{4p}^{(n+1)}
            &\gets \min_{\bm{\alpha}_{4p}}
            E_{\text{atom}}^{\text{MP2,corr}}(\bm{\alpha}_{\text{val}}^{(n+1)}, \bm{\alpha}_{4p})
            + \frac{\varepsilon_{0,4p}}{\kappa_0} \times
            \max_{M \in \{M\}} \kappa(
                \bm{\alpha}_{\text{val}}^{(n+1)},
                \bm{\alpha}_{4p},
                \bm{\alpha}_{\text{pol}}^{(n)}; M
            )   \\
        \bm{\alpha}_{\text{pol}}^{(n+1)}
            &\gets \min_{\bm{\alpha}_{\text{pol}}}
            E_{\text{atom}}^{\text{MP2,corr}}(
                \bm{\alpha}_{\text{val}}^{(n+1)},
                \bm{\alpha}_{4p}^{(n+1)},
                \bm{\alpha}_{\text{pol}}
            )
            + \frac{\varepsilon_{0,\text{pol}}}{\kappa_0} \times
            \max_{M \in \{M\}} \kappa(
                \bm{\alpha}_{\text{val}}^{(n+1)},
                \bm{\alpha}_{4p}^{(n+1)},
                \bm{\alpha}_{\text{pol}}; M
            )
    \end{split}
    \end{equation}
    Because the $\kappa$ penalty couples all three exponent subsets, the optimizations in \cref{eq:mcao_loss_three_term} are iterated to convergence. Here, $\bm{\alpha}^{(n)}$ denotes the exponents at the $n$th macrocycle.
    We use three macrocycles for all MCAO optimizations, which are sufficient to converge the atomic energy to reasonable precision.
    The $\varepsilon_0$ parameter is set to $1$, $0.1$, and $0.01$~m$E_{\text{h}}$ for the valence, $4p$, and polarization optimizations, respectively.

    \begin{figure}[!h]
        \centering
        \includegraphics[width=3in]{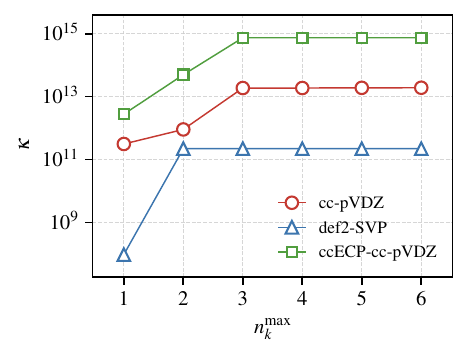}
        \caption{Convergence of the condition number of various basis sets with respect to the $k$-mesh size for \textit{fcc} Cu.
        }
        \label{fig:Cu_cond_kconv}
    \end{figure}

    For a given reference material, the condition number $\kappa$ is defined as
    \begin{equation}
        \kappa_{\mathcal{K}}
            = \frac{
                \max_{\bm{k} \in \mathcal{K}} \max_{i} |\lambda_{\bm{k}i}|
            }{
                \min_{\bm{k} \in \mathcal{K}} \min_{i} |\lambda_{\bm{k}i}|
            }
    \end{equation}
    where $\lambda_{\bm{k}i}$ are the eigenvalues of the basis-set overlap matrix evaluated at $k$-point $\bm{k}$ for the reference solid.
    Thus, $\kappa$ depends on the $k$-point set $\mathcal{K}$ used to sample the first Brillouin zone.
    In our implementation, we first determine the maximum $k$-mesh dimensions along the three reciprocal-lattice directions, $(N_{1}^{\text{max}}, N_{2}^{\text{max}}, N_{3}^{\text{max}})$, such that the corresponding $N_{1}^{\text{max}} \times N_{2}^{\text{max}} \times N_{3}^{\text{max}}$ Born--von Karman supercell is close to cubic and contains more than $300$ atoms, i.e.,
    \begin{equation}
        N_{1}^{\text{max}} N_{2}^{\text{max}} N_{3}^{\text{max}} n_{\text{atom}} \geq 300
    \end{equation}
    where $n_{\text{atom}}$ is the number of atoms per unit cell.
    For \textit{fcc} Cu, $n_{\text{atom}} = 4$ and we set $N_{1}^{\text{max}} = N_{2}^{\text{max}} = N_{3}^{\text{max}} = n_{k}^{\text{max}}$ for its cubic symmetry.
    We then construct $\mathcal{K}$ from all symmetrically unique $k$-points in $\Gamma$-centered meshes of size $N_{1} \times N_{2} \times N_{3}$, where $1 \leq N_{\alpha} \leq N_{\alpha}^{\text{max}}$ for $\alpha = 1, 2, 3$.
    \Cref{fig:Cu_cond_kconv} shows that for \textit{fcc} Cu, $\kappa$ converges very rapidly and $n_{k}^{\text{max}} = 3$ essentially saturates $\kappa$.
    Throughout this work, we use a relatively conservative value of $n_{k}^{\text{max}} = 5$.

\begin{figure}[!h]
        \centering
        \includegraphics[width=6in]{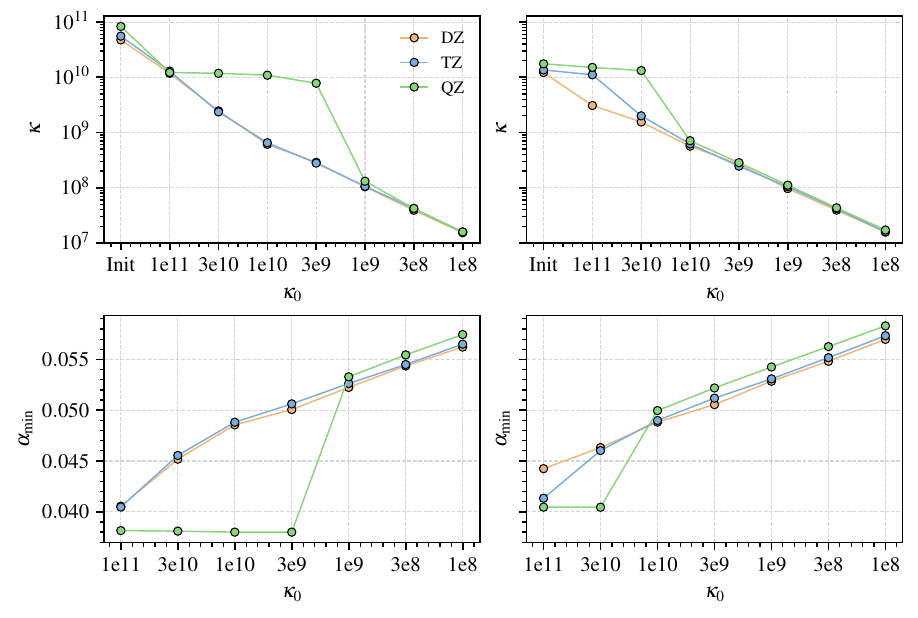}
        \caption{Variation of the condition number (upper row) and minimum primitive exponent (bottom row) as a function of the applied $\kappa_0$ parameter during MCAO.
        The left two panels are results for the nonrelativistic all-electron basis sets, while the right two panels are results for the SFX2C-1e all-electron basis sets.
        In all cases, the minimum exponent comes from the $s$-channel.
        }
        \label{fig:cond_exp_evolve}
    \end{figure}

    To avoid numerical problems caused by an excessively large $\kappa$ penalty, we adjust $\kappa_0$ dynamically during MCAO.
    In particular, we select initial basis sets whose condition numbers are within $10^{2}$ of the final value of $\kappa_0$.
    In this work, the final value of $\kappa_0$ is $10^{9}$, so the initial basis sets have condition numbers below $10^{11}$.
    MCAO then starts with $\kappa_0 = 10^{11}$, so that the $\kappa$ penalty is at most $\varepsilon_0$ in magnitude.
    We progressively decrease $\kappa_0$ by half a logarithmic unit, i.e.,~$10^{10.5}, 10^{10}, \cdots$, until reaching the final target $\kappa_0 = 10^{9}$.
    At each $\kappa_0$, we perform three macro iterations of \cref{eq:mcao_loss_three_term} as discussed above and use the optimized exponents as input for the next $\kappa_0$ value.
    This procedure gives a relatively smooth decrease in $\kappa$ for the optimized basis sets, as shown in the upper row of \cref{fig:cond_exp_evolve}.
    Some nonsmooth behavior can occur in the early stages, when $\kappa$ is still high, likely because the optimization is more susceptible to local minima.
    The reduction in $\kappa$ is driven by a continuous increase in the exponents of the most diffuse primitive functions, as shown in the bottom row of \cref{fig:cond_exp_evolve}.

    \begin{table}[!h]
        \centering
        \caption{Valence-set structures (primitive and contracted) used in this work for different nuclear-potential and relativistic treatments.
        The minimum valence exponents from $s$, $p$, and $d$ channels and the condition number of the final MCAO-cc-pV$X$Z basis sets (including the polarization set) evaluated on \textit{fcc} Cu are also shown for comparison.
        }
        \label{tab:basis_struct}
        \begin{tabular}{lcccccccc}
            \toprule
            & Zeta & Primitive & Contracted & $N_{\text{AO}}$
                & $\alpha_{\text{min}}^{s}$
                & $\alpha_{\text{min}}^{p}$
                & $\alpha_{\text{min}}^{d}$
                & $\kappa$  \\
            \midrule
            \multirow{3}*{GTH-PBE-lc} & DZ & $3s(2+1)p3d$ & $2s2p2d$ & $25$ & $0.048$ & $0.224$ & $0.401$ & $4.57 \times 10^{4}$  \\
            & TZ & $3s(2+1)p4d$ & $3s3p3d$ & $50$ & $0.052$ & $0.207$ & $0.222$ & $8.73 \times 10^{7}$  \\
            & QZ & $3s(2+1)p4d$ & $3s3p4d$ & $82$ & $0.052$ & $0.207$ & $0.222$ & $2.29 \times 10^{8}$  \\
            &  &  &  &  &  &  &  &   \\
            \multirow{3}*{GTH-PBE-sc} & DZ & $5s(4+1)p4d$ & $3s2p2d$ & $26$ & $0.047$ & $0.194$ & $0.308$ & $6.33 \times 10^{5}$  \\
            & TZ & $5s(4+1)p5d$ & $4s3p3d$ & $51$ & $0.051$ & $0.187$ & $0.200$ & $1.65 \times 10^{8}$  \\
            & QZ & $5s(4+1)p5d$ & $5s4p4d$ & $87$ & $0.051$ & $0.187$ & $0.200$ & $6.74 \times 10^{8}$  \\
            &  &  &  &  &  &  &  &   \\
            \multirow{3}*{ccECP-soft} & DZ & $6s(4+1)p4d$ & $3s2p2d$ & $26$ & $0.047$ & $0.201$ & $0.334$ & $1.72 \times 10^{6}$  \\
            & TZ & $6s(4+1)p5d$ & $4s3p3d$ & $51$ & $0.050$ & $0.192$ & $0.219$ & $1.91 \times 10^{8}$  \\
            & QZ & $6s(4+1)p5d$ & $5s4p4d$ & $87$ & $0.050$ & $0.192$ & $0.219$ & $6.21 \times 10^{8}$  \\
            &  &  &  &  &  &  &  &   \\
            \multirow{3}*{ccECP-hard} & DZ & $6s(5+1)p5d$ & $3s2p2d$ & $26$ & $0.046$ & $0.196$ & $0.344$ & $1.40 \times 10^{6}$  \\
            & TZ & $6s(5+1)p6d$ & $4s3p3d$ & $51$ & $0.049$ & $0.187$ & $0.239$ & $1.74 \times 10^{8}$  \\
            & QZ & $6s(5+1)p6d$ & $5s4p4d$ & $87$ & $0.049$ & $0.187$ & $0.239$ & $6.16 \times 10^{8}$  \\
            &  &  &  &  &  &  &  &   \\
            \multirow{3}*{All-e/NR} & DZ & $16s(10+1)p5d$ & $5s3p2d$ & $31$ & $0.048$ & $0.197$ & $0.389$ & $1.38 \times 10^{6}$  \\
            & TZ & $16s(10+1)p7d$ & $6s4p3d$ & $56$ & $0.053$ & $0.184$ & $0.221$ & $8.46 \times 10^{7}$  \\
            & QZ & $16s(10+1)p7d$ & $7s5p4d$ & $92$ & $0.053$ & $0.185$ & $0.225$ & $1.23 \times 10^{8}$  \\
            &  &  &  &  &  &  &  &   \\
            \multirow{3}*{All-e/SFX2C-1e} & DZ & $17s(10+1)p5d$ & $5s3p2d$ & $31$ & $0.049$ & $0.207$ & $0.385$ & $1.79 \times 10^{6}$  \\
            & TZ & $17s(10+1)p7d$ & $6s4p3d$ & $56$ & $0.053$ & $0.193$ & $0.219$ & $8.44 \times 10^{7}$  \\
            & QZ & $17s(10+1)p7d$ & $7s5p4d$ & $92$ & $0.054$ & $0.193$ & $0.218$ & $1.04 \times 10^{8}$  \\
            \bottomrule
        \end{tabular}
    \end{table}

    \subsection{Valence set size selection}
    \label{subsec:val_select}

    The valence-set sizes used in this work for different nuclear potentials and zeta levels are summarized in \cref{tab:basis_struct}.
    \myred{For the TZ and QZ valence sets, we choose the number of primitives in each angular-momentum channel to reduce the atomic HF energy error to approximately $1$~milli-Hartree.
    When this produces overly diffuse primitives, we reduce their number in the affected channel until the initial basis satisfies $\kappa < 10^{11}$.
    For DZ, we use a smaller $d$ shell matching def2-SVP to obtain more systematic zeta convergence.}
    % For all nuclear potentials, the TZ and QZ basis sets share the same valence primitive set (but contracted differently), which we found to improve extrapolation to the CBS limit.
    % For the DZ valence set, using the same valence primitive set as in TZ and QZ leads to larger RPA@PBE errors for bulk Cu.
    % We therefore vary the size of the $d$ shell and choose the DZ valence set that gives the smallest RPA@PBE error in bulk Cu benchmarks relative to the extrapolated CBS(TZ,QZ) reference.
    % For all nuclear-potential and relativistic treatments considered here, this selection gives a DZ valence set with a slightly smaller $d$ shell than the corresponding TZ and QZ sets.
    \Cref{tab:basis_struct} also lists the minimum primitive exponent in each valence angular-momentum channel and the condition number of the full basis set, including the polarization functions, evaluated for \textit{fcc} Cu.
    As discussed in the main text, the selected valence-set sizes retain similar diffuse primitives in each angular-momentum channel, giving comparable condition numbers for the reference solid.

    \section{Auxiliary basis sets}
    \label{sec:aux_basis}

    For each AO basis set, we optimize an auxiliary basis for accurate and efficient density-fitting calculations.
    The auxiliary basis is even-tempered and fully uncontracted, with Gaussian exponents in each angular-momentum channel given by
    \begin{equation}
        \alpha_{l,i}
            = \alpha_{l} \beta_{l}^{i-1},
        \qquad{}
        i = 1, 2, \cdots, N_l
    \end{equation}
    The parameters $\bm{p} = \{\alpha_l, \beta_l\}_{l=0}^{l_{\text{max}}}$ are determined by minimizing a loss function that combines HF and MP2 fitting errors,
    \begin{equation}    \label{eq:aux_loss}
        \mathcal{L}_{\text{aux}}(\bm{p})
            = \max \{
                \gamma \Delta \mathbf{J}^{\text{HF}},
                \gamma \Delta \mathbf{K}^{\text{HF}},
                \Delta E_{J}^{\text{HF}},
                \Delta E_{K}^{\text{HF}},
                \Delta E_{\text{corr}}^{\text{MP2}},
                \Delta \mathbf{T}^{\text{MP2}}_2
            \}.
    \end{equation}
    Here,
    \begin{itemize}
        \item $\Delta \mathbf{J}^{\text{HF}}$ and $\Delta \mathbf{K}^{\text{HF}}$ are the maximum absolute errors of the HF Coulomb and exchange matrices,
        \item $\Delta E_{J}$ and $\Delta E_{K}$ are the absolute errors of the HF Coulomb and exchange energies,
        \item $\Delta E_{\text{corr}}^{\text{MP2}}$ is the absolute error of the MP2 correlation energy,
        \item and $\Delta \mathbf{T}^{\text{MP2}}_2$ is the maximum absolute error of the MP2 $T_2$ amplitudes.
    \end{itemize}
    All terms in \cref{eq:aux_loss} are evaluated in atomic units for a single atom (for Cu, we use the $[\text{Ar}]3d^{10}4s^{1}$ configuration).
    Because commonly used auxiliary basis sets often have relatively large $\Delta \mathbf{J}^{\text{HF}}$ and $\Delta \mathbf{K}^{\text{HF}}$, we set $\gamma = 0.1$ to allow a looser tolerance for these two matrix errors.

    Minimizing the auxiliary loss in \cref{eq:aux_loss} determines the optimal parameters for an auxiliary basis set of fixed size.
    For practical density-fitting calculations, the auxiliary basis should be as compact as possible while maintaining the required fitting precision.
    We therefore start from a sufficiently large set whose optimized loss is below $\tau_{\text{aux}} = 10^{-5}$ and prune it iteratively until the optimized loss exceeds $\tau_{\text{aux}}$.
    At each iteration, we consider removing one primitive from each angular-momentum channel in turn. We reoptimize the remaining exponents in that channel while keeping the parameters of all other channels fixed.
    This produces a pool of candidate sets, each with one fewer primitive in a single angular-momentum channel.
    The candidate with the lowest optimized loss is accepted, and the parameters of the full set are reoptimized before the next pruning iteration.

    \begin{figure}[!h]
        \centering
        \includegraphics[width=6in]{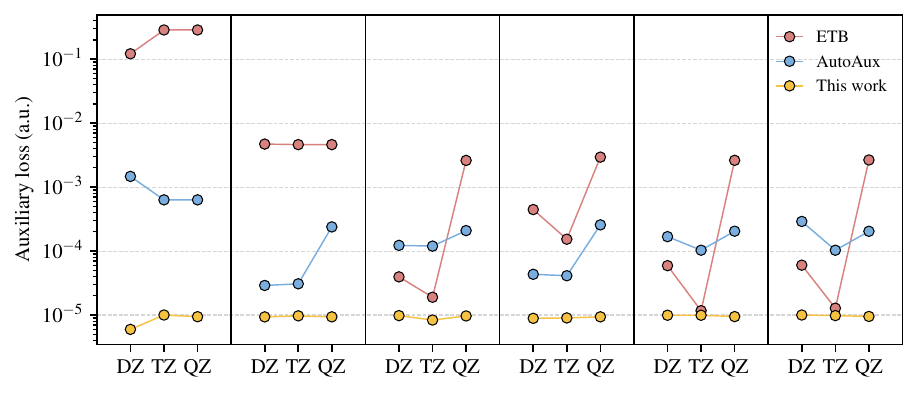}
        \caption{Auxiliary loss defined in \cref{eq:aux_loss} for different auxiliary basis sets on a single Cu atom.
        From left to right, the nuclear-potential and relativistic treatments are GTH-PBE-lc, GTH-PBE-sc, ccECP-hard, ccECP-soft, all-e/NR, and all-e/SFX2C-1e.
        }
        \label{fig:auxloss}
    \end{figure}

    \begin{figure}[!h]
        \centering
        \includegraphics[width=6in]{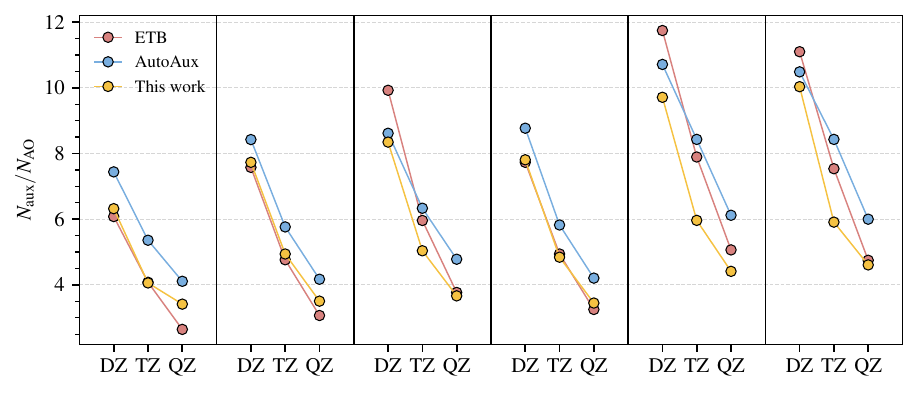}
        \caption{Auxiliary-basis-set sizes for a single Cu atom, expressed as ratios to the corresponding AO basis-set sizes.
        From left to right, the nuclear-potential and relativistic treatments are GTH-PBE-lc, GTH-PBE-sc, ccECP-hard, ccECP-soft, all-e/NR, and all-e/SFX2C-1e.
        }
        \label{fig:auxrat}
    \end{figure}

    % Autoaux: https://pubs.acs.org/doi/10.1021/acs.jctc.6b01041

    \Cref{fig:auxloss} shows the loss for auxiliary basis sets optimized for MCAO-cc-pV$X$Z with different pseudopotentials, ECPs, and all-electron potentials.
    The optimized auxiliary bases achieve the target loss of $10^{-5}$~a.u.~for all nuclear-potential treatments and AO $\zeta$ levels.
    For comparison, we also include two commonly used automatic choices: an even-tempered basis (ETB) with $\beta = 2$ and the AutoAux basis developed by Neese and co-workers~\cite{Stoychev17JCTC}, both generated using their PySCF implementations.
    ETB is the default auxiliary basis in PySCF when a named auxiliary basis is unavailable for a given element, while AutoAux is the default in ORCA~\cite{Neese12WIRCMS}.
    As shown in \cref{fig:auxloss}, these automatic auxiliary bases either have errors substantially above $10^{-5}$~a.u.~or show variable precision across $\zeta$ levels and nuclear potentials.
    The optimized auxiliary bases are also more compact than the automatic alternatives in almost all cases, as shown in \cref{fig:auxrat}.

    \section{\textit{fcc} Bulk Copper}
    \label{sec:bulk}

    \begin{figure}[!h]
        \centering
        \includegraphics[width=6in]{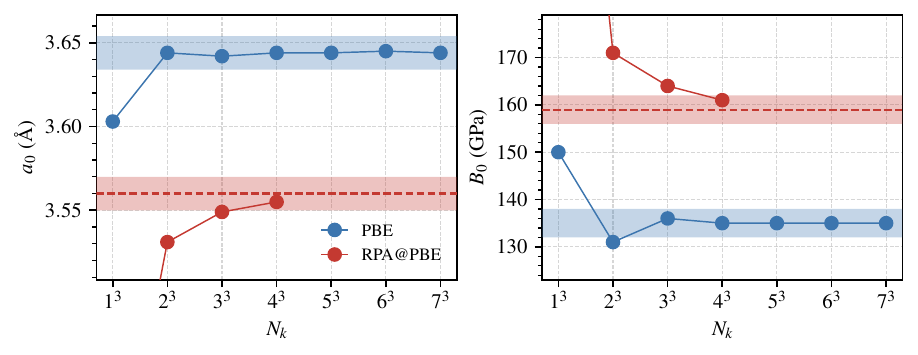}
        \caption{Convergence of the PBE and RPA@PBE lattice constant ($a_0$) and bulk modulus ($B_0$) for \textit{fcc} Cu using the MCAO-cc-pVDZ basis set optimized for the GTH-PBE-sc pseudopotential.
        For PBE, the blue shaded area indicates deviations within $0.01~\text{\AA}$ in $a_0$ and $3$~GPa in $B_0$ from the $N_k = 7^3$ results.
        For RPA@PBE, the red shaded area indicates deviations within $0.01~\text{\AA}$ in $a_0$ and $3$~GPa in $B_0$ from the TDL-extrapolated results.
        }
        \label{fig:bulk_a0b0_conv}
    \end{figure}

    The lattice constant $a_0$ and bulk modulus $B_0$ are obtained by fitting single-point energies calculated at lattice constants near the experimental value to the Birch--Murnaghan equation of state.
    For the conventional \textit{fcc} cell (four Cu atoms per cell) used in this work, the Brillouin zone is sampled by $n \times n \times n$ evenly spaced $k$-point meshes of size $N_k = n^3$.
    These $k$-meshes are shifted by the Baldereschi mean-value point~\cite{Baldereschi73PRB} (MVP)
    \begin{equation}    \label{eq:bulk_mvp}
        \bm{k}_{\text{MVP}}^{\textit{fcc}}
            = (0.6223, 0.2953, 0)
    \end{equation}
    to reduce finite-size errors (FSEs).
    \Cref{fig:bulk_a0b0_conv} shows the convergence of both $a_0$ and $B_0$ with $N_k$ for PBE and RPA@PBE, evaluated using the MCAO-cc-pVDZ basis set with the GTH-PBE-sc pseudopotential.
    For PBE, $N_k = 4^3$ converges $a_0$ within $0.01~\text{\AA}$ and $B_0$ within $1$~GPa relative to the $N_k = 7^3$ reference.
    We therefore use $N_k = 4^3$ without further extrapolation.
    RPA@PBE converges more slowly than PBE with respect to $N_k$.
    We obtain the TDL estimates of $a_0$ and $B_0$ by first extrapolating the total RPA@PBE energies from $N_k = 3^3$ and $N_k = 4^3$ to the TDL using the $O(N_k^{-1})$ FSE scaling~\cite{Xing24PRX} and then fitting the resulting energy curve to the Birch--Murnaghan equation of state.
    In \cref{fig:bulk_a0b0_conv}, the extrapolated RPA@PBE $a_0$ and $B_0$ are shown as horizontal dashed lines.
    The largest finite-$k$-mesh results from $N_k = 4^3$ are within $0.01~\text{\AA}$ and $3$~GPa of the TDL-extrapolated values, supporting the reliability of the extrapolation.

    After obtaining TDL energies in the DZ basis set, we apply the following composite correction to reach the combined CBS and thermodynamic limit for both PBE and RPA@PBE:
    \begin{equation}
        E(\text{CBS}, \text{TDL})
            = E(\text{DZ}, \text{TDL})
            + E(\text{CBS(T,Q)}, N_k^*)
            - E(\text{DZ}, N_k^*)
    \end{equation}
    where $N_k^* = 2^3$ with the same Baldereschi MVP shift.
    For the PBE energy and the mean-field part of the RPA energy, we extrapolate to the CBS limit using the exponential formula from ref~\citenum{Feller93JCP,Jensen05TCA,Karton06TCA,Varandas19PCCP}:
    \begin{equation}    \label{eq:cbs_mf}
        E(\text{CBS})
            = \frac{
                \exp(\beta X_1) E(X_1) - \exp(\beta X_2) E(X_2)
            }{
                \exp(\beta X_1) - \exp(\beta X_2)
            }
    \end{equation}
    Here, $\beta = 1.54$, as suggested by Karton and Martin~\cite{Karton06TCA}. For the RPA correlation energy, we use the standard Helgaker $X^{-3}$ formula~\cite{Helgaker97JCP}:
    \begin{equation}    \label{eq:cbs_corr}
        E(\text{CBS})
            = \frac{
                X_1^{3} E(X_1) - X_2^{3} E(X_2)
            }{
                X_1^{3} - X_2^{3}
            }
    \end{equation}
    In both cases, $X_1 = 3$ and $X_2 = 4$ for TZ and QZ, respectively.
    The final CBS and TDL values of $a_0$ and $B_0$ are listed in \cref{tab:final_a0b0}.

    \begin{table}[!h]
        \centering
        \caption{Final $a_0$ (in $\text{\AA}$) and $B_0$ (in GPa) calculated using different nuclear-potential and relativistic treatments, extrapolated to the CBS limit using the MCAO-cc-pV(T,Q)Z basis sets.}
        \label{tab:final_a0b0}
        \begin{tabular}{lcccc}
            \toprule
            \multirow{2}*{Method}
                & \multicolumn{2}{c}{PBE}
                & \multicolumn{2}{c}{RPA@PBE}   \\
            \cmidrule(lr){2-3} \cmidrule(lr){4-5}
                & $a_0$ & $B_0$
                & $a_0$ & $B_0$ \\
            \midrule
            GTH-PBE-lc & $3.647$ & $134$ & $3.508$ & $218$  \\
            GTH-PBE-sc & $3.629$ & $141$ & $3.592$ & $152$  \\
            ccECP-soft & $3.631$ & $137$ & $3.588$ & $148$  \\
            ccECP-hard & $3.649$ & $133$ & $3.603$ & $145$  \\
            All-e/NR & $3.663$ & $127$ & $3.632$ & $132$  \\
            All-e/SFX2C-1e & $3.623$ & $142$ & $3.591$ & $147$  \\
            \bottomrule
        \end{tabular}
    \end{table}

    For the TZV2P-MOLOPT and def2-SVP results shown in Fig.~4 of the main text, we apply only the TDL extrapolation described above; no basis-set extrapolation is performed.

    For the PBE band structure, we first perform an SCF calculation using a $3\times3\times3$ $k$-point mesh shifted by the Baldereschi MVP. We then calculate the bands along the chosen path without further SCF iterations.

    \section{CO adsorption on Cu(111)}
    \label{sec:surf}

    \subsection{Final $E_{\text{ads}}$}

    \Cref{tab:final_eads} lists the final adsorption energies for the top and hollow sites, converged to the combined basis-set and thermodynamic limit using the protocol detailed below.
    The surface model and geometry protocol largely follow ref~\citenum{Cao25arXiv}; the principal differences are our approach to TDL convergence and the use of the MCAO basis sets developed here.

    \begin{table}[!h]
        \centering
        \caption{Final adsorption energies (eV) of CO/Cu(111).
        The RPA@PBE values are the numbers reported in Fig.~5 of the main text.}
        \label{tab:final_eads}
        \begin{tabular}{lcccccc}
            \toprule
            \multirow{2}*{Method} & \multicolumn{3}{c}{PBE} & \multicolumn{3}{c}{RPA@PBE}  \\
            \cmidrule(lr){2-4}\cmidrule(lr){5-7}
            & Top & Hollow & Difference
            & Top & Hollow & Difference \\
            \midrule
            GTH-PBE-lc & $-0.73$ & $-0.86$ & $-0.12$ & $-0.77$ & $-0.75$ & $0.01$  \\
            GTH-PBE-sc & $-0.77$ & $-0.91$ & $-0.14$ & $-0.37$ & $-0.25$ & $0.12$  \\
            ccECP-soft & $-0.76$ & $-0.89$ & $-0.13$ & $-0.35$ & $-0.24$ & $0.11$  \\
            ccECP-hard & $-0.71$ & $-0.83$ & $-0.12$ & $-0.30$ & $-0.18$ & $0.12$  \\
            All-e/NR & $-0.72$ & $-0.85$ & $-0.12$ & $-0.25$ & $-0.13$ & $0.12$  \\
            All-e/SFX2C-1e & $-0.79$ & $-0.94$ & $-0.14$ & $-0.34$ & $-0.24$ & $0.10$  \\
            & & & & & & \\
            PAW/PW (this work) & $-0.78$ & $-0.91$ & $-0.13$ & & & \\
            \bottomrule
        \end{tabular}
    \end{table}

    \subsection{Geometry}

    Following previous work~\cite{Cao25arXiv}, the Cu(111) surface is simulated using a slab model with a $2\times2$ supercell in the $xy$ plane, four atomic layers, and $15~\text{\AA}$ of vacuum in the $z$ direction.
    We construct the surface using the bulk PBE lattice constant $a_0 = 3.634~\text{\AA}$.
    We then place a single CO molecule at either the top or \textit{fcc} hollow site and relax the geometry in VASP using an unshifted $6\times6\times1$ $k$-point mesh.
    The bottom two layers are fixed during the geometry relaxation, while all other atoms are fully relaxed.
    To remain consistent with previous work~\cite{Ren09PRB,Schimka10NM}, we do not use dispersion corrections in the geometry relaxations or subsequent PBE single-point energy calculations.

    \subsection{Adsorption energy}

    Following ref~\cite{Ye24FD,Cao25arXiv}, the adsorption energy is calculated as
    \begin{equation}    \label{eq:Eads}
        E_{\text{ads}}
            = E_{\text{surf+mol}} - E_{\text{surf}} - E_{\text{mol}}
            \approx E_{\text{int}} + \Delta_{\text{geom}}
    \end{equation}
    where $E_{\text{surf+mol}}$, $E_{\text{surf}}$, and $E_{\text{mol}}$ are the energies of the adsorption system, the free surface, and the free molecule, respectively.
    We separate the adsorption energy into two parts: $E_{\text{int}}$ is the adiabatic interaction energy calculated with the surface and molecule frozen at their adsorption geometries, and $\Delta_{\text{geom}}$ is the energy change associated with relaxation of the isolated fragments.
    This decomposition has two practical advantages.
    \begin{itemize}
        \item First, basis-set superposition error (BSSE) can be corrected using the standard counterpoise (CP) approach~\cite{Boys70MP} in the interaction-energy calculation,
        \begin{equation}
            E_{\text{int}}
                = E_{\text{surf+mol}} - E_{\text{surf@surf+mol}} - E_{\text{mol@surf+mol}}
        \end{equation}
        where $E_{\text{surf@surf+mol}}$ and $E_{\text{mol@surf+mol}}$ are the energies of the surface and molecule, respectively, evaluated at their frozen adsorption geometries in the combined surface+molecule basis set.

        \item Second, previous work shows that surface relaxation is negligible~\cite{Neef06SS,Cao25arXiv}, so $\Delta_{\text{geom}}$ includes only relaxation of the CO molecule and is evaluated at the PBE level, i.e.,
        \begin{equation}    \label{eq:delta_geom}
            \Delta_{\text{geom}}
                = E^{\text{PBE}}_{\text{CO(ads)}} - E^{\text{PBE}}_{\text{CO(g)}}
        \end{equation}
        where CO(ads) and CO(g) denote CO in the adsorption and gas-phase geometries, respectively.
        The CO geometry is characterized by its bond length.
        We use bond lengths of $1.157~\text{\AA}$, $1.181~\text{\AA}$, and $1.135~\text{\AA}$ for the top-site, hollow-site, and gas-phase geometries, respectively, in close agreement with ref~\citenum{Cao25arXiv}.
        The final relaxation energies extrapolated to the CBS limit using TZ and QZ basis sets for different nuclear-potential and relativistic treatments are listed in \cref{tab:surf_relax}.
    \end{itemize}

    \begin{table}[!h]
        \centering
        \caption{CO relaxation energy (eV) defined in \cref{eq:delta_geom}, evaluated at PBE level and extrapolated to the CBS limit using TZ and QZ basis sets.
        }
        \label{tab:surf_relax}
        \begin{tabular}{lcc}
            \toprule
            Method & Top & Hollow \\
            \midrule
            GTH-PBE  & $0.022$ & $0.099$ \\
            ccECP    & $0.034$ & $0.123$ \\
            All-e/NR & $0.023$ & $0.102$ \\
            All-e/SFX2C-1e & $0.023$ & $0.102$ \\
            \bottomrule
        \end{tabular}
    \end{table}

    \subsection{Converging $E_{\text{int}}$ to the combined basis-set and thermodynamic limit}

\begin{figure}[!h]
        \centering
        \includegraphics[width=6.5in]{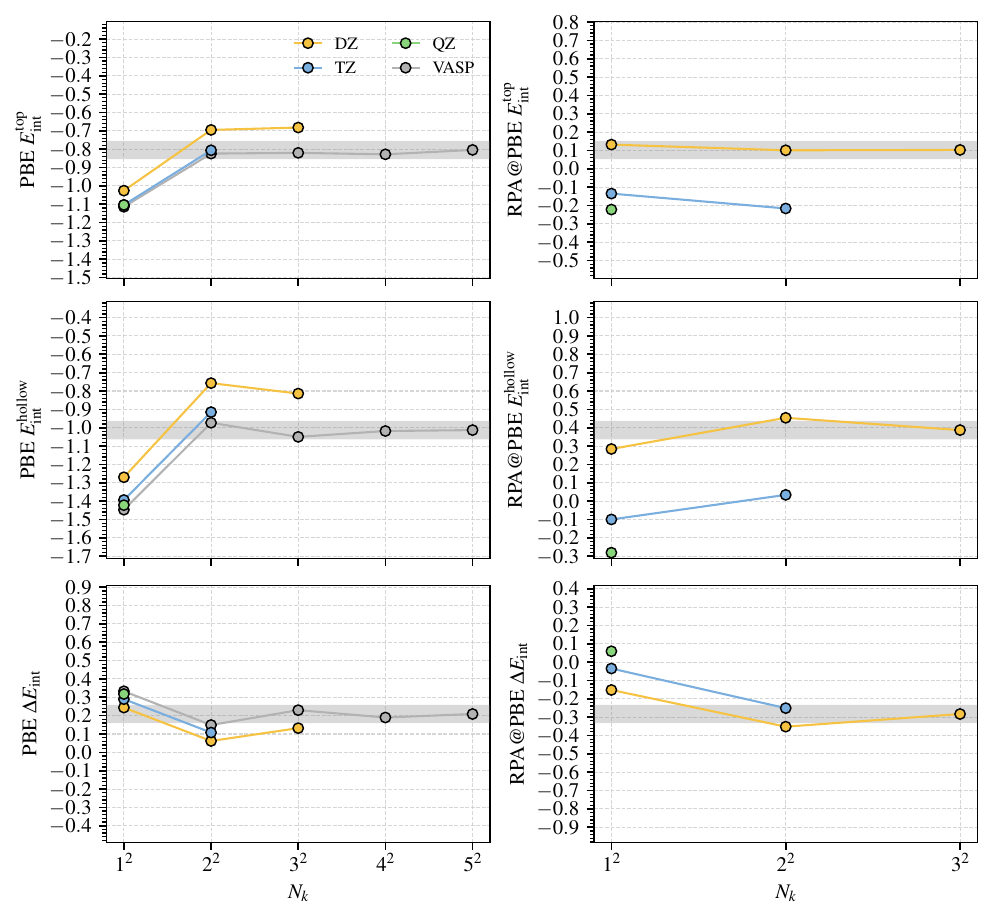}
        \caption{Convergence of the PBE and RPA@PBE interaction energies with $k$-mesh size, obtained using the MCAO-cc-pV$X$Z basis sets with the GTH-PBE-sc pseudopotential.
        VASP PAW/PW results for PBE, obtained using the structures optimized in this work, are included for comparison. The gray shaded area indicates deviations within $\pm 0.05$~eV from the VASP/$5\times5$ results.
        For RPA@PBE, the gray shaded area indicates deviations smaller than $\pm 0.05$~eV from the DZ/$3\times3$ results, which are the largest-$k$-mesh RPA calculations performed here.
        }
        \label{fig:eads_conv}
    \end{figure}

    To establish an efficient protocol with controlled basis-set and finite-size errors, we calculated PBE and RPA@PBE $E_{\text{int}}$ using several combinations of basis sets and $k$-point meshes with the GTH-PBE-sc pseudopotential. The results are summarized in \cref{fig:eads_conv}.
    As in the bulk calculations, shifting the $k$-point meshes by an MVP improves the convergence of $E_{\text{int}}$ with $k$-mesh size.
    We tested two MVPs.
    The first is the exact Baldereschi MVP for a two-dimensional hexagonal lattice~\cite{Baldereschi73PRB},
    \begin{equation}
        \bm{k}_{\text{MVP}}^{\textit{hex}}
            = (0.7614, 0.3807, 0)
    \end{equation}
    and the second is an approximate Baldereschi-style MVP derived by requiring only the first-shell Fourier component to vanish for a two-dimensional hexagonal lattice,
    \begin{equation}    \label{eq:surf_mvp}
        \bm{k}_{\text{MVP}}^{\textit{hex}\text{(1)}}
            = (0.6667, 0.6667, 0).
    \end{equation}
    Both MVPs give faster convergence than $\Gamma$-centered meshes; we use \cref{eq:surf_mvp} for the final calculations because it performs slightly better for our slab model.

    We also performed VASP PBE calculations to study FSE decay using larger $k$-meshes than are feasible with Gaussian basis sets.
    The VASP PBE $E_{\text{int}}$ values converge rapidly with $k$-mesh size; the $3\times3$ mesh is already accurate to better than $0.05$~eV.
    The Gaussian-basis results obtained with the MCAO-cc-pV$X$Z sets follow the same convergence pattern as the VASP results, although the maximum $k$-mesh size accessible in PySCF is more limited.
    We therefore use the $3\times3$ $k$-mesh for the DZ calculations and apply composite corrections at smaller $k$-meshes to remove the remaining basis-set incompleteness error.
    Our final protocol for PBE $E_{\text{int}}$ is as follows:
    \begin{equation}    \label{eq:e_int_cbs_tdl}
    \begin{split}
        E_{\text{int}}(\text{CBS}, \text{TDL})
            &= E_{\text{int}}(\text{DZ}, 3\times3)    \\
            &\quad{}+ E_{\text{int}}(\text{TZ}, 2\times2)
            - E_{\text{int}}(\text{DZ}, 2\times2)    \\
            &\quad{}+ E_{\text{int}}(\text{CBS(T,Q)}, 1\times1)
            - E_{\text{int}}(\text{TZ}, 1\times1)
    \end{split}
    \end{equation}
    where the CBS(T,Q) extrapolation is performed using \cref{eq:cbs_mf}.

    For RPA@PBE, although no PW reference is available, the $3\times3$ mesh also converges the DZ $E_{\text{int}}$ to within approximately $0.05$~eV, as indicated by the small difference between the $2\times2$ and $3\times3$ results.
    Moreover, the DZ and higher-zeta basis sets show nearly parallel $k$-mesh convergence, supporting the use of composite corrections at smaller $k$-meshes for the remaining basis-set incompleteness error.
    We therefore apply the same protocol in \cref{eq:e_int_cbs_tdl} to the RPA@PBE calculations.
    In this case, the CBS(T,Q) extrapolation is performed separately for the HF@PBE energy and the RPA correlation energy using \cref{eq:cbs_mf,eq:cbs_corr}, respectively.

    \subsection{Results from def2 basis sets}

    Following ref~\citenum{Cao25arXiv}, we also calculated all-electron nonrelativistic $E_{\text{ads}}$ using modified def2-SVP and def2-TZVP basis sets with primitives below $0.05$~a.u.~removed.
    To focus on basis-set effects, we use the same PBE-optimized surface models as in the MCAO calculations and approach the combined basis-set and thermodynamic limit using a protocol inspired by \cref{eq:e_int_cbs_tdl},
    \begin{equation}    \label{eq:e_int_cbs_tdl_dztz}
    \begin{split}
        E_{\text{int}}(\text{CBS}, \text{TDL})
            &= E_{\text{int}}(\text{DZ}, 3\times3)    \\
            &\quad{}+ E_{\text{int}}(\text{CBS(D,T)}, 2\times2)
            - E_{\text{int}}(\text{DZ}, 2\times2)
    \end{split}
    \end{equation}
    where CBS(D,T) is extrapolated using \cref{eq:cbs_mf,eq:cbs_corr} with $X_1 = 2$ for def2-SVP and $X_2 = 3$ for def2-TZVP.
    Using alternative two-point CBS extrapolation formulas does not change the qualitative conclusions below.

    The results are listed in \cref{tab:dztz_eads}.
    As noted in the main text, these values differ from those reported in ref~\citenum{Cao25arXiv} with the same def2-(S,TZ)VP basis sets, likely because of differences in the treatment of finite-size effects.
    Compared with the basis-set-converged all-e/NR results in \cref{tab:final_eads}, the PBE $E_{\text{ads}}$ values from the def2-(S,TZ)VP basis sets are reasonable, with errors within $0.05$~eV.
    In contrast, the RPA results have sizable, site-dependent basis-set errors, shifting the predicted site-preference energy by about $0.2$~eV.
    At least part of this error arises from the absence of the QZ-quality information used in the CBS(T,Q) extrapolation.
    Consistent with this interpretation, \cref{tab:dztz_eads} also includes results obtained using \cref{eq:e_int_cbs_tdl_dztz} with our MCAO-cc-pV(D,T)Z basis sets, which are in reasonable agreement with the def2-(S,TZ)VP results.

    \begin{table}[!h]
        \centering
        \caption{All-electron nonrelativistic PBE and RPA@PBE $E_{\text{ads}}$ (eV) calculated using \cref{eq:e_int_cbs_tdl_dztz} for the def2-(S,TZ)VP series and the MCAO-cc-pV(D,T)Z series.
        }
        \label{tab:dztz_eads}
        \begin{tabular}{lcccccc}
            \toprule
            \multirow{2}*{Method} & \multicolumn{3}{c}{PBE} & \multicolumn{3}{c}{RPA@PBE}  \\
            \cmidrule(lr){2-4}\cmidrule(lr){5-7}
            & Top & Hollow & Difference
            & Top & Hollow & Difference \\
            \midrule
            def2-(S,TZ)VP & $-0.68$ & $-0.78$ & $-0.10$ & $0.08$ & $0.38$ & $0.29$  \\
            MCAO-cc-pV(D,T)Z & $-0.63$ & $-0.71$ & $-0.08$ & $0.13$ & $0.44$ & $0.31$  \\
            \bottomrule
        \end{tabular}
    \end{table}

    \section{Addressing linear dependence using canonical orthogonalization}

    \begin{figure}[!h]
        \centering
        \includegraphics[width=6.5in]{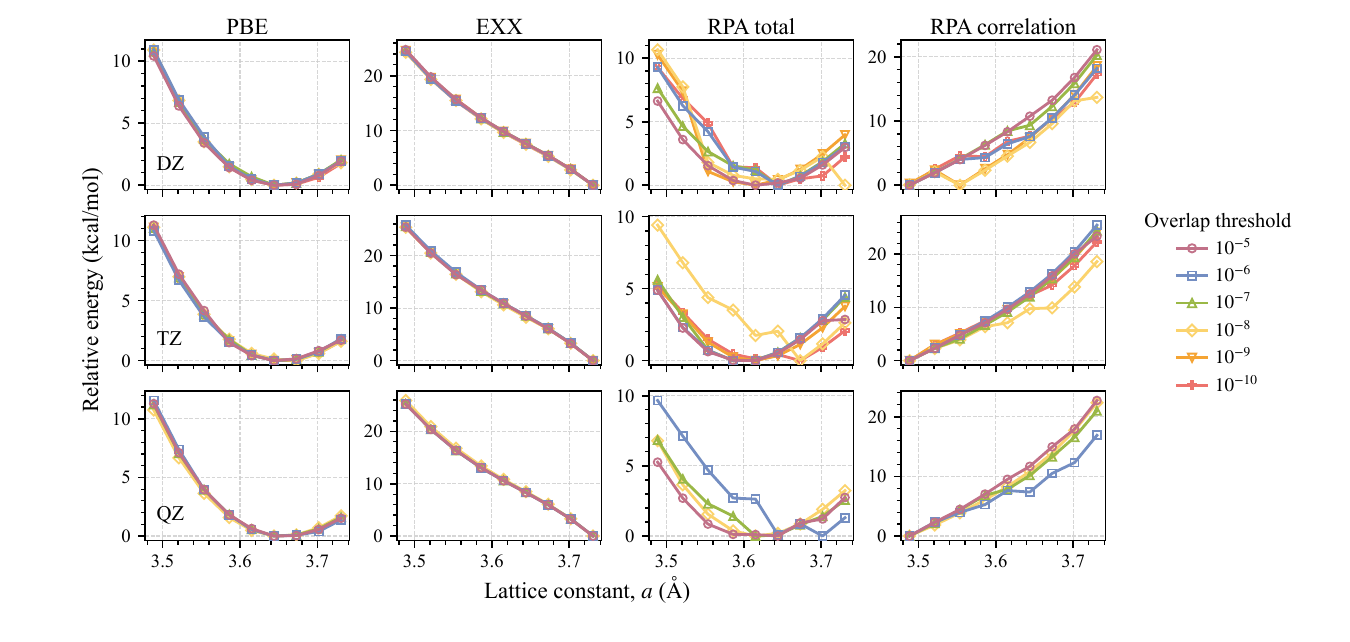}
        \caption{Energy--lattice constant curves for \textit{fcc} Cu obtained with atom-optimized cc-pV$X$Z basis sets. Shown are the PBE energy, the HF energy evaluated using PBE orbitals (EXX), the RPA correlation energy evaluated using PBE orbitals, and the RPA total energy, which is the sum of EXX and RPA correlation energies.
        Curves obtained with different numerical thresholds for canonical orthogonalization are shown.
        The PBE and EXX curves remain smooth over the lattice constants considered, whereas the RPA correlation energy is more sensitive to the threshold.
        Consequently, the RPA total-energy curves show kinks in most cases.
        }
        \label{fig:cu_eos}
    \end{figure}

    \begin{figure}[!h]
        \centering
        \includegraphics[width=5in]{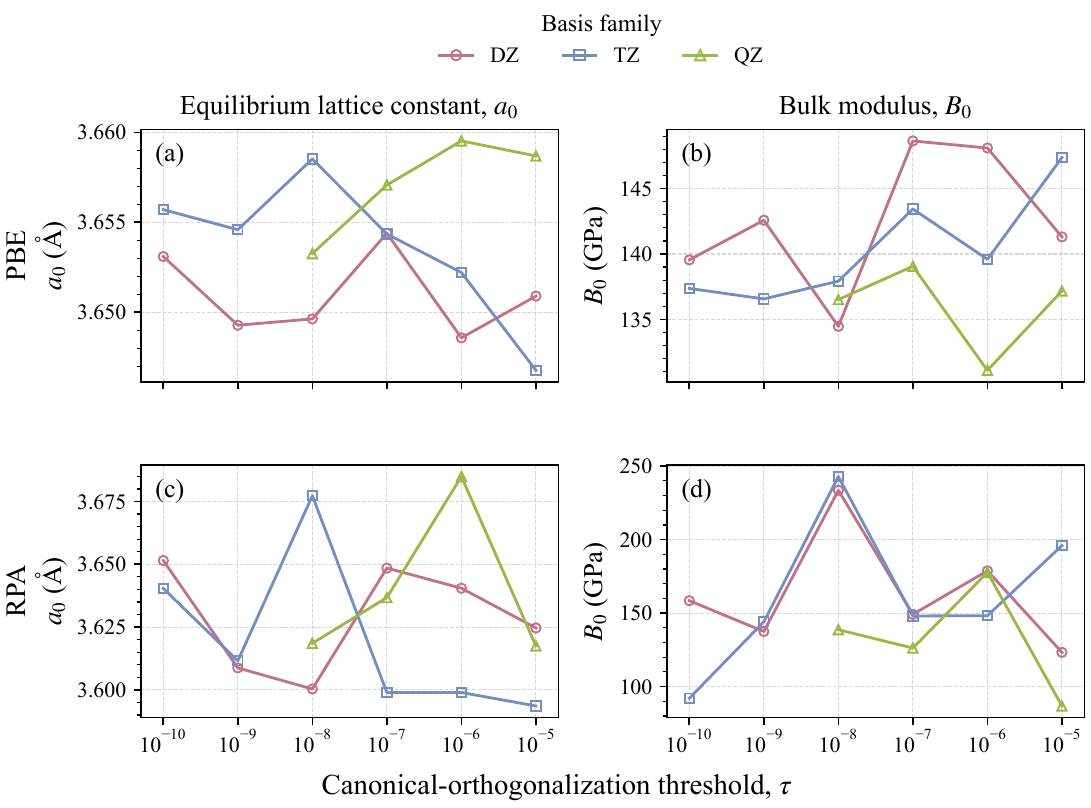}
        \caption{Lattice constants ($a_0$) and bulk moduli ($B_0$) obtained by fitting the PBE and RPA total-energy data in \cref{fig:cu_eos} to the Birch--Murnaghan equation of state for different numerical thresholds in canonical orthogonalization.
        The fitted $a_0$ and $B_0$ show moderate fluctuations at the PBE level but much larger fluctuations at the RPA level, preventing their accurate extraction from RPA calculations with canonical orthogonalization.
        }
        \label{fig:cu_eos_fit}
    \end{figure}

    \clearpage

    \bibliography{refs_si}